\documentclass[aps,prl,superscriptaddress,showpacs,amsmath,amssymb,nofootinbib,reprint]{revtex4-2}
\usepackage{amsmath}
\usepackage{amsfonts}
\usepackage{amsthm}
\usepackage{graphics}
\usepackage{graphicx}
\usepackage{amssymb}
\usepackage{mathtools}
\usepackage{color}
\usepackage{url}
\usepackage[abs]{overpic}
\usepackage[usenames,dvipsnames]{xcolor}
\usepackage[normalem]{ulem}
\usepackage{braket}
﻿

﻿

﻿
      
﻿

﻿
﻿
\def\da{\downarrow}
\def\ua{\uparrow}

﻿
﻿
\begin{document}
﻿
\title{Controlling quantum scars and engineering subharmonic responses with a two frequency drive}
﻿
\author{Pinaki Dutta}
\affiliation{Department of Physics, Indian Institute of Technology Kharagpur, Kharagpur - 721 302, India}
﻿
\author{Kamal L Panigrahi}
\affiliation{Department of Physics, Indian Institute of Technology Kharagpur, Kharagpur - 721 302, India}

\author{Vishwanath Shukla}
\email{vishwanath.shukla@phy.iitkgp.ac.in}
\affiliation{Department of Physics, Indian Institute of Technology Kharagpur, Kharagpur - 721 302, India}%
﻿
\date{\today}
\begin{abstract} 
We demonstrate that a continuous two frequency drive is a versatile and robust protocol to control the lifetime of quantum many body scars and to engineer non-equilibrium phases of driven quantum matter. By modulating the frequency ratio $c$ (any rational number), we systematically explore prethermal features across a broad frequency range. For small integer values of $c$, we observe ergodicity breaking even at moderately low frequencies, signaling long-lived scarred dynamics. By continuously increasing $c$, one can generate non-monotonic transitions between ergodic and non-ergodic dynamics. These observations are consistent with the predictions of an effective Floquet Hamiltonian based approach. Furthermore, we exploit this tunability to engineer fractional subharmonic responses, highlighting the potential of two-frequency driving as a theoretical platform for controlling scars, prethermalization, and time crystal-like behavior.
\end{abstract}
﻿
﻿
\maketitle
﻿
\paragraph*{Introduction}

Floquet engineering, over the past two decades, has emerged as a promising avenue for controlling and manipulating quantum matter. From tuning transport properties to realizing artificial Hamiltonians and exotic non-equilibrium phases, such as time crystals, periodic modulation offers a comparatively richer platform than static systems~\cite{lindner2011floquet, kundu2014effective, else2016floquet, aditya2023dynamical, zhang2017observation, gangopadhay2025counterdiabatic, guo2025dynamical, oka2019floquet, bai2023floquet, cayssol2013floquet, rudner2020band, sacha2017time}. The heating of such systems to a featureless infinite temperature state, predicted by the Floquet Eigenstate Thermalization Hypothesis (ETH)~\cite{srednicki1994chaos, d2016quantum, mori2018thermalization, lazarides2014equilibrium, d2013many, d2014long}, is often mitigated by generating high-frequency prethermal states, wherein the energy absorption from the drive is effectively suppressed~\cite{ho2023quantum, canovi2016stroboscopic, abanin2017effective, bukov2015universal, bukov2016heating, abanin2015exponentially}.

A systematic control of the prethermal features is also essential for the advancement of quantum technologies. As a generalization of single-frequency drives, there is a growing interest in implementing multiple-frequency schemes~\cite{ikeda2022floquet, pena2024steering, kar2017two, kar2016tuning, else2020long, das2023periodically, mori2021rigorous, kumar2024prethermalization, yan2024prethermalization}, which have the potential to realize nontrivial topological phases~\cite{martin2017topological, qi2024real,long2022many, long2021nonadiabatic, crowley2019topological}. The presence of multiple harmonics maps the problem onto a multidimensional Floquet lattice, consisting of the system’s spatial dimensions and the synthetic frequency dimensions, thereby enabling the modeling of topological properties of higher-dimensional systems~\cite{lin2016photonic, ozawa2016synthetic, yuan2016photonic}. Moreover, the mitigation of the heating problem has also been addressed in the Hubbard model using such drives~\cite{murakami2023suppression,chen2024mitigating}. Recent studies have shown that a two-frequency modulation is more effective in protecting superconducting qubits from a low-frequency noise and enhancing coherence by engineering dynamical sweet spots~\cite{briseno2025dynamical, hajati2024dynamic}.

Several mechanisms, such as many-body localization~\cite{nandkishore2015many, alet2018many, altman2018many, lazarides2015fate, abanin2019colloquium}, quantum many-body scars (QMBS)~\cite{serbyn2021quantum, turner2018weak, bernien2017probing, mcclarty2020disorder, mohapatra2023pronounced, chandran2023quantum, zhao2020quantum}, and Hilbert-space fragmentation~\cite{moudgalya2022quantum,moudgalya2022hilbert}, can result in either weak or strong violations of thermalization. A paradigmatic model for realizing quantum scars is the PXP model, where scarred eigenstates emerge due to an approximate SU(2) algebra~\cite{choi2019emergent}. These scarred states lead to a long-lived coherent dynamics and are expected to protect quantum information~\cite{zhang2023many, serbyn2021quantum}, particularly in quantum sensing tasks~\cite{dooley2021robust, dooley2023entanglement, desaules2022extensive}, where entanglement can easily introduce decoherence. The Floquet version of the PXP model has been extensively studied to understand the fate of scars, including their role in entanglement control, time-crystal formation, and scar enhancement~\cite{hudomal2022driving, maskara2021discrete, bluvstein2021controlling, mukherjee2020collapse, mukherjee2022periodically, huang2024engineering, dutta2025prethermalization}. Also, a two-rate protocol has previously been used to demonstrate the existence of Floquet bands~\cite{banerjee2024exact} and heating suppression in the PXP model~\cite{ghosh2025heating}.

In this study, we analytically and numerically illustrate that a continuous two-frequency drive, characterized by frequencies $\omega_1$ and $\omega_2 = c\,\omega_1$ with $c$ a rational number, provides a generic mechanism to control the lifetime of scar-induced oscillations. In particular, when $c$ is tuned to a small integer, the scarring features can persist down to much lower frequencies even at strong driving, thereby suppressing the heating or entanglement growth and enabling the prethermalization. This behaviour has direct implications for the frequency of a revival, exhibiting a subharmonic response, allowing it to be regulated as a function of $c$. Also, we observe several instances of fractional response. Interestingly, small values of $c$ support stable higher-order subharmonic responses, an effect that is absent for the single-frequency drive. We support our arguments by computing the return probability, entanglement entropy and magnetization.

\begin{figure*}
	\includegraphics[width=0.95\linewidth]{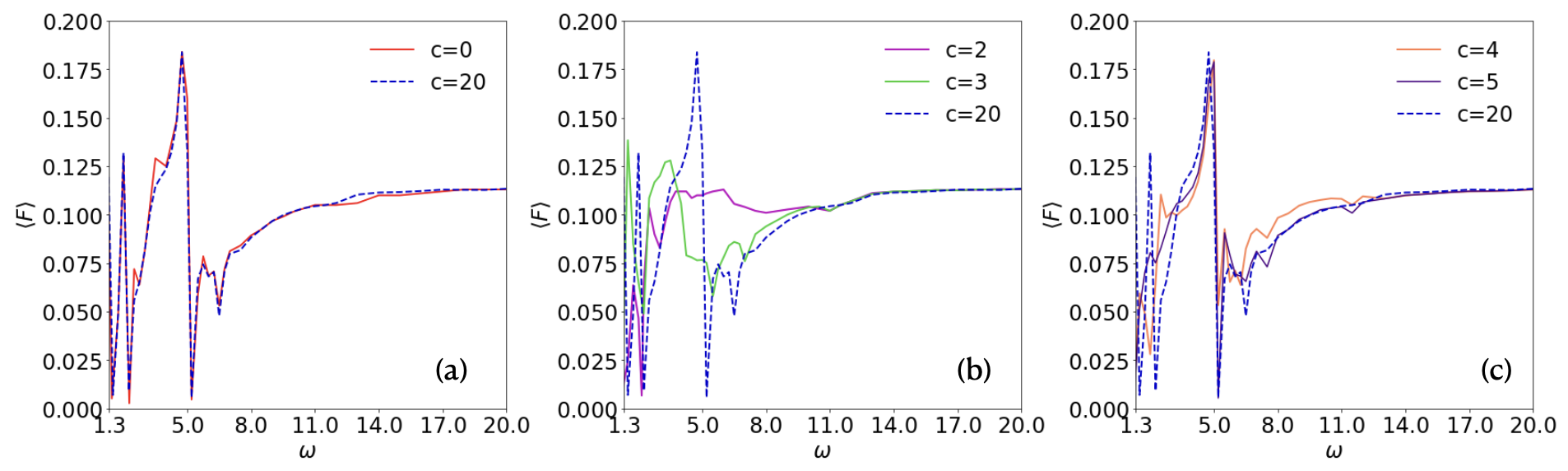}
	\caption{\textbf{Integral values of the ratio $\omega_2/\omega_1$}. Time-averaged $\langle F \rangle$ vs $\omega$ for integral values of the ratio $c=\omega_2/\omega_1$. (a) The equivalence between a single-frequency $c = 0$ and a two-frequency $c = 20$  drives is demonstrated, both exhibit multiple non-monotonic transitions consistent with analytical predictions. (b) and (c) When $c$ is tuned to $2$, $3$ and $4$, the high-frequency transitions vanish, $\langle F \rangle$ remains non-zero for $\omega \geq 2.14$. However, at $c = 5$, a non-monotonic transition emerges for $\omega \approx 5.15$. Independent of $c$, $\langle F \rangle$ saturates to approximately $0.11$ at sufficiently high frequencies. $\lambda_0 = 12$ is held fixed for all the cases.
	}
	\label{fig:1}
\end{figure*}

\paragraph*{Theoretical framework and numerical simulations}
We consider a kinetically constrained chain of spin-$1/2$ particles described by the PXP Hamiltonian
\begin{equation} \label{eq:PXPmodel}
	H = - \Omega \sum_i P_{i-1} \sigma^x_i P_{i+1} + \frac{\lambda}{2}\sum_i \sigma^z_i,
\end{equation}
where $\sigma^{x,z}_i$ is a Pauli matrix at site $i$, the projector $P_i=(1-\sigma^z_i)/2$ imposes a constraint that two neighboring sites cannot be simultaneously in the excited (up) state. We define $P_{i-1} \sigma^x_i P_{i+1} \equiv \tilde{\sigma}^x_i$ to rewrite the Hamiltonian as $H=-\Omega \sum_i \Tilde{\sigma}^x_i + \frac{\lambda}{2}\sum_i \sigma^z_i$ and drive the chain using a two frequency continuous periodic drive
\begin{equation}
	\lambda(t) = \lambda_0 [\sin\omega_1 t + \sin \omega_2 t ],
\end{equation}
where $\lambda_0$ is the drive amplitude, $\omega_1=\omega$ and $\omega_2=c\omega$ are the drive frequencies, with $c$ a rational number. 
﻿

To gain insights into the stroboscopic dynamics, we use the effective Hamiltonian, $H_F$, defined by the Floquet theorem for the evolution operators $U=\exp(-i\int^T_0 H(t)\, dt) \equiv \exp(-i H_F T)$, along with Floquet perturbation theory (FPT) by rewriting the Hamiltonian as $H=H_0(t) + V$, where $H_0(t) = (\lambda_0/2)[\sin \omega t + \sin c\omega t]$ and treat $V=-\Omega \sum_i \tilde{\sigma}^i_x$ as a perturbation; $T$ is the drive period. FPT remains applicable at any driving frequency and is particularly well-suited for the high amplitude regime ($\lambda_0 \gg \Omega$). $H_F$ so obtained depends on the nature of the ratio $c$, whether it is an integer or a fraction.

﻿

We use the above prescription to obtain the first order Hamiltonian $H_F(1)=H^{(a)}_F+H^{(b)}_F$ for the integral values of $c$, where
\begin{flalign*}
	H_{F}^{(a)} & = -J_{0}\left(\frac{\lambda_0}{\omega}\right) J_{0}\left(\frac{\lambda_0}{c \omega}\right)  
	\biggl[ \cos\left( \frac{(c+1)\lambda_0}{c\omega} \right) \sum_{j} \tilde{\sigma}_{j}^{x}  \\
	& \quad - \sin\left( \frac{(c+1)\lambda_0}{c\omega} \right) \sum_{j} \tilde{\sigma}_{j}^{y} \biggr],
\end{flalign*}
\begin{flalign*}
	& H_{F}^{(b)}  = -\Biggl[ e^{i\frac{(c+1)\lambda_0}{c\omega}} \sum_{\beta\neq0} i^{\beta(1-c)} J_{-c\beta} \left(-\frac{\lambda_0}{\omega}\right) J_{\beta} \left(-\frac{\lambda_0}{c\omega}\right) \\ 
	& \times \sum_{j} \tilde{\sigma}_{j}^{+} 
	 + e^{-i\frac{(c+1)\lambda_0}{c\omega}} \sum_{\beta\neq0} i^{\beta(1-c)} J_{-c\beta} \left(\frac{\lambda_0}{\omega}\right) J_{\beta} \left(\frac{\lambda_0}{c\omega}\right) \\
	 & \quad \quad \times  \sum_{j} \tilde{\sigma}_{j}^{-} \Biggr],
\end{flalign*}
$J_0(x)$ is the Bessel function of zeroth order, and $\sigma^x = \tilde{\sigma}^+ + \tilde{\sigma}^-$, with the action defined on the eigenstates $\ket{n}$ of $S^z=\sum_i \sigma^z_i$ as $\tilde{\sigma}^{\pm}\ket{n}=\ket{n\pm 1}$. The calculation of higher order terms is cumbersome, but we find that $H_F(2)=0$ and $H_F(3) \sim O(\Omega^3) \sum_i \tilde{\sigma}^+_{i-1} \tilde{\sigma}^+_{i+1}\tilde{\sigma}^-_i$, indicating the presence of a longer range non-PXP terms.
﻿

However, when $c=p/q$, a fraction, where $p$ and $q$ are integers and $q\neq 1$, $H^{(b)}_F$ is modified to 
\begin{multline}
H_{F}^{(b)} = -[
e^{i \left(1+\frac{q}{p}\right) \frac{\lambda_0}{\omega}} 
\sum_{\beta \neq 0} i^{\beta(q-p)} 
J_{-p\beta} \left(-\frac{\lambda_0}{\omega}\right) \\
\times J_{q \beta} \left(-\frac{\lambda_0}{c\omega}\right) 
 \sum_{j} \tilde{\sigma}_{j}^{+} 
+ e^{-i \left(1+\frac{q}{p}\right) \frac{\lambda_0}{\omega}} 
\sum_{\beta \neq 0} i^{\beta(q-p)} 
J_{-p\beta} \left(\frac{\lambda_0}{\omega}\right)\\ 
\times
J_{q \beta} \left(\frac{\lambda_0}{c\omega}\right)
\sum_{j} \tilde{\sigma}_{j}^{-}],
\end{multline}
whereas $H^{(a)}_F$ remains identical to the integral $c$ case.
﻿

In this study, we set $L=22$, $\Omega=1$ and use periodic boundary conditions, see Supplemental Material  for system size dependence~\cite{supplemental}. We initiate the nonequilibrium dynamics from the N\'eel state $\ket{Z_2}=\ket{\ua \da \ua ..}$, unless otherwise stated. We characterize the dynamics by computing the fidelity (return probability) $F(t)=|\braket{\psi(t)|\psi(0)}|^2$ and its time-average over the period $\tau$ as $\langle F \rangle = (1/\tau) \int^{\tau}_0 F(t) \, dt$.
﻿

﻿
\begin{figure*}
	\includegraphics[width=0.95\linewidth]{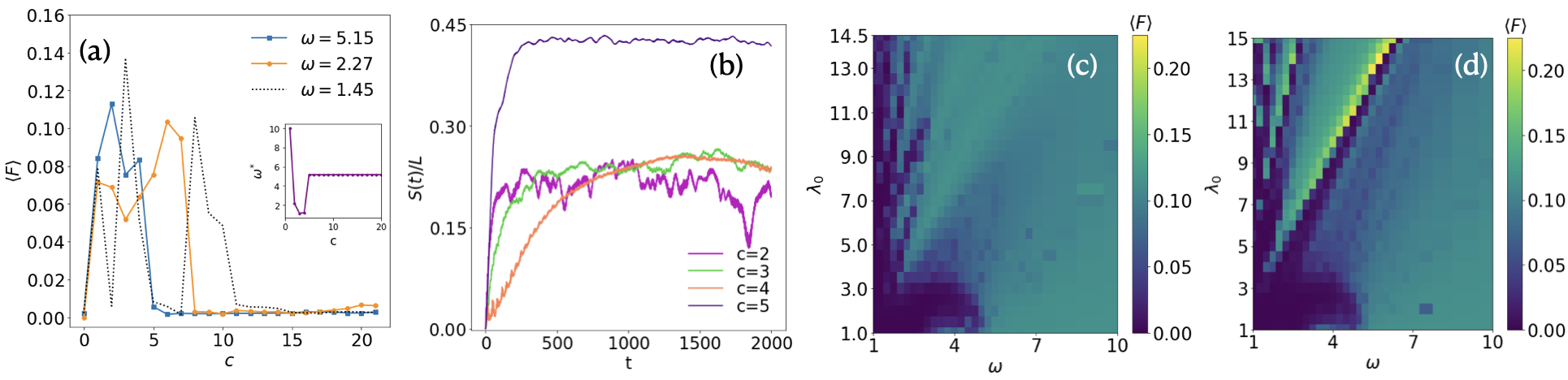}
	\caption{(a) Time-averaged fidelity $\langle F \rangle$ vs integer values of $c$, near the first three zeros of $J_0(\lambda_0/\omega)$. Inset: Critical frequency $\omega^{*}$, the point at which ergodicity first emerges, vs $c$. (b) Entanglement entropy at $\omega \approx 5.15$ for $c = 2, \,3$ and $4$ exhibits a suppressed growth, whereas saturates to a ETH value for $c = 5$. In (a) and (b) $\lambda_0$ is set to 12. (c) and (d): Density plots of $\langle F \rangle$ for $c = 2$ and $c = 20$, respectively.}
	\label{fig:2}
\end{figure*}
﻿
\begin{figure}
	\includegraphics[width=0.95\linewidth]{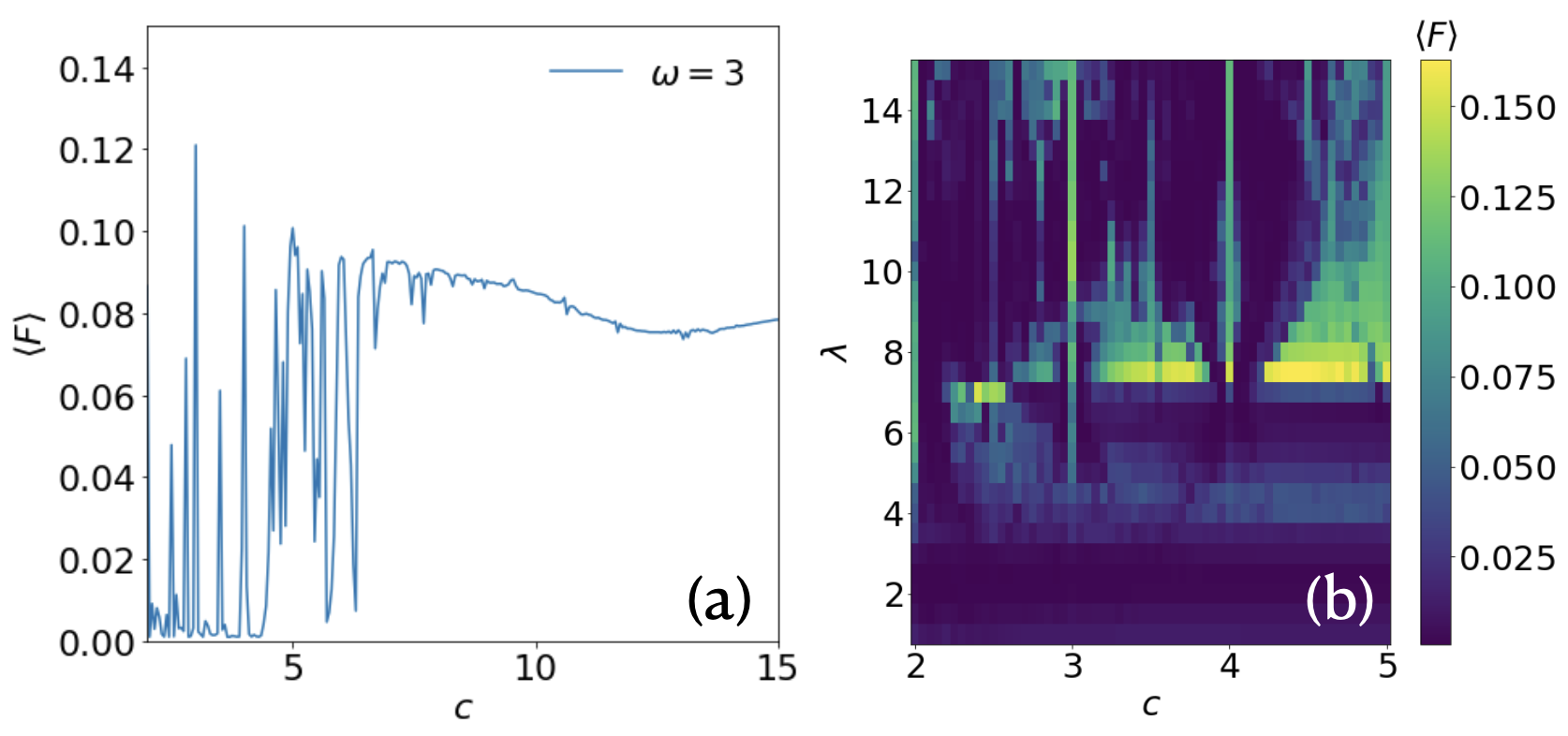}
	\caption{\textbf{Fractional values of the ratio $\omega_2/\omega_1$}.  
(a) Time-averaged fidelity $\langle F \rangle$ vs $c$, with increase in $c$ the non-monotonic transitions gradually disappear; $\langle F \rangle$ saturates to a finite value for $\omega=3$ and $\lambda_0=12$.  (b) Density plot of $\langle F \rangle$ illustrates the dependence of these transitions on $\lambda_0$ for $\omega=3$.
}
	\label{fig:3}
\end{figure}
﻿
\begin{figure}
	\includegraphics[width=0.95\linewidth]{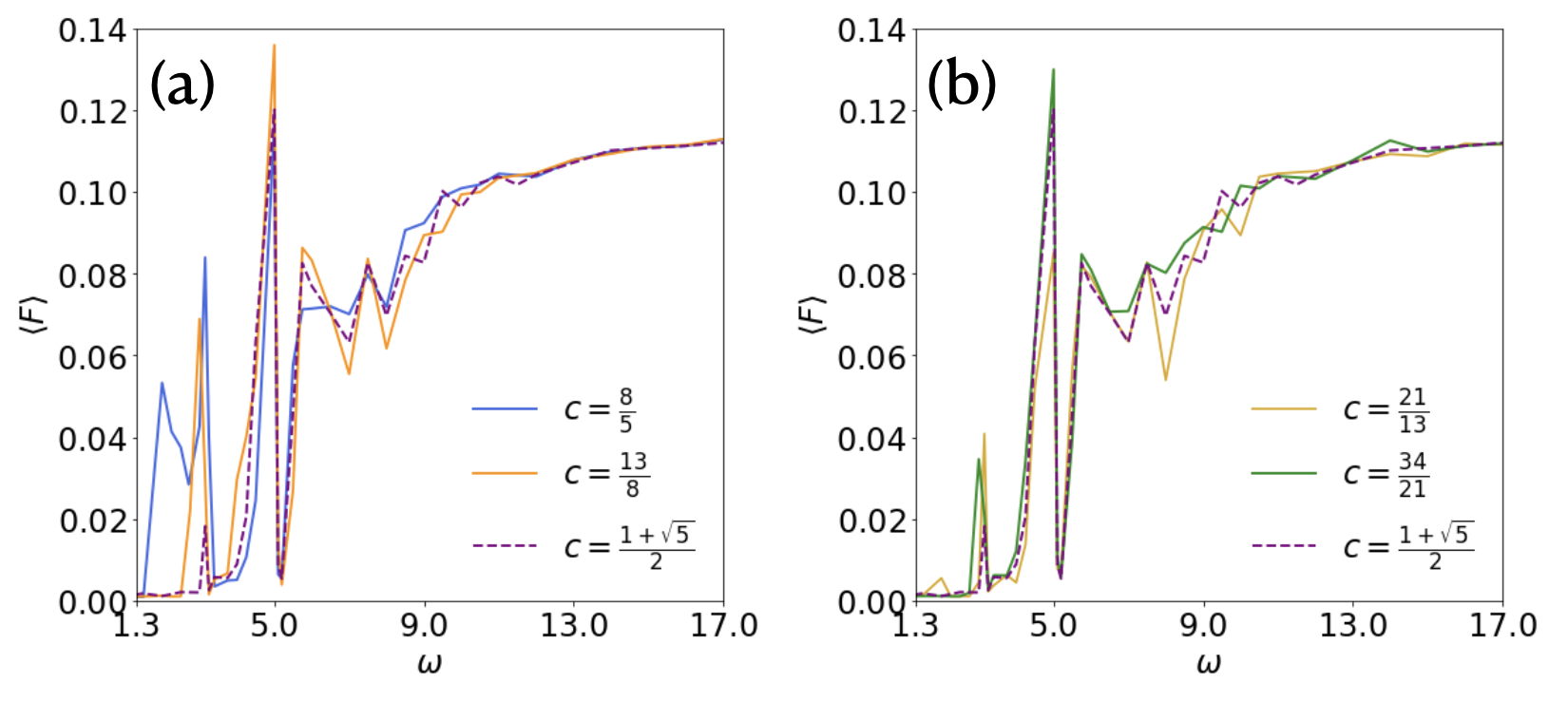}
	\caption{\textbf{Approximating a quasi-periodic drive}.  (a) $c=8/5$ and $13/8$. (b) $c=21/8$ and $34/21$. The dynamics gradually approaches that of the exact quasi-periodic case as $c$ approaches the value $(1+\sqrt{5})/2$. $\lambda_0 = 12$ is kept fixed.}
	\label{fig:4}
\end{figure}
﻿
﻿
\paragraph*{Integral values of the ratio $\omega_2/\omega_1$} 
Figure~\ref{fig:1} (a) shows that for a large integer value of $c=20$ and $\omega < 5.5$, $\langle F \rangle$ exhibits several non-monotonic transitions between ergodic ($\langle F \rangle \sim 0$) and non-ergodic ($\langle F \rangle \neq 0$) regimes. The first order Floquet hamiltonian $H_F(1)$ vanishes in the limit $c \gg 1$, when the frequencies are tuned to result in $J_0(\lambda_0/\omega) =0$, making it consistent with single-frequency drive. Thus, for $\lambda_0/\omega \approx \, 2.40, \, 5.5, \, 8.65, \hdots$,  $H^{(a)}_F = 0 $ and $H^{(b)}_F \approx 0$, as $J_{-cn}(\lambda_0/\omega) \approx 0$. Also, the dynamics is fully controlled by the non-PXP terms in $H_F(3)$.

﻿
For $\omega \geq 5.5$, $\langle F \rangle$ increases monotonically, indicating the presence of scar-induced oscillations and eventually saturates to $\sim 0.1$. Also, in the limit $\omega \gg \lambda_0$, $J_0(\lambda_0/\omega) \to 1$ and $J_n(\lambda_0/\omega) \to 0$, Eq.~\eqref{eq:PXPmodel} reduces to the static PXP hamiltonian, which exhibits QMBS. However, for smaller values of $c$, both $J_0(\lambda_0/\omega)$ and $J_{-cn}(\lambda_0/\omega)$ may not simultaneously drop to zero. Hence, the characteristics of return probability may differ significantly from those observed for a single-frequency drive. We illustrate this for $c=2$ and $c=3$ in Fig.~\ref{fig:1} (b). Even though at sufficiently high frequencies scar-induced oscillations are present, indicating an absence of dependence on $c$, but as we decrease the frequency, these oscillations persist until at very low frequencies ($\omega_{*} \approx 2.13$ for $c=2$ and $\omega_{*} \approx 1.2$ for $c=3$). This behavior signals the onset of ergodic dynamics, which is consistent with the ETH. Similarly, for $c=4$, $\langle F \rangle$ drops to zero only around $\omega \approx 1.1$, whereas for $c=5$ we observe a non-monotonic behavior near $\omega \approx 5.15$, which coincides with one of the zeros of $J_0(\lambda_0/\omega)$, see Fig.~\ref{fig:1} (c). Thus, non-monotonic transitions can be suppressed by tuning to smaller $c$ values. Also, we can obtain scar-controlled prethermal features over a wide range of $\omega$ values.

﻿
The above features prompted us to systematically examine the dependence of dynamics on $c$ at the special frequencies $\omega = 1.45, \, 2.27$ and $5.15$, which we show in Fig.~\ref{fig:2} (a). We find that $\langle F \rangle$ does not vanish simultaneously for these frequencies. For $\omega = 5.15$, $\langle F \rangle \approx 0$ when $c$ exceed a threshold value, $c \geq 5$. The threshold increases to $8$ for $\omega = 2.27$. However, when we decrease the frequency to $\omega=1.45$, $\langle F \rangle$ shows a complex behavior, but for $c > 10$ the dynamics essentially becomes ergodic, resembling that of a single-frequency drive.

﻿
In the inset of Fig.~\ref{fig:2} (a), we show the variation with $c$ of the critical frequency $\omega^*$ around which $\langle F \rangle$ drops to zero for the first time (the onset of ergodicity), while coming down from the high frequency side. Moreover, in Fig.~\ref{fig:2} (b) we illustrate the behavior at $\omega=5.15$, when $c$ is varied, by observing the dynamical evolution of the half-chain entanglement entropy $S_{E}(t)=T_{R}[\rho_R \log \rho_R]$, where the reduced density matrix for the right half of the chain $\rho_R(t)=T_L[\ket{\psi(t)}\bra{\psi(t)}]$ is obtained by tracing over the degrees of freedom of the left half of the chain. For $c=5$, $S_E(t)$ grows rapidly and settles to a value indistinguishable from the infinite temperature ETH value, while it grows relatively slowly and saturates to a much lower value for $c<5$. Therefore, by tuning the ratio of the two driving frequencies, we can effectively control the lifetime of scars. 
﻿

Finally, we examine the influence of the driving strength $\lambda_0$ in Fig.~\ref{fig:2} (c) and (d), by comparing the density plots of $\langle F \rangle$ in the $\lambda_0-\omega$ plane for $c=2$ and $20$. In the low-frequency regime ($\omega < 2$), the dynamics is predominantly ergodic with respect to variations in $\lambda_0$. However, at large $\lambda_0 > 6$, the dynamics does not exhibit non-monotonic transitions in a significant manner as we vary $\omega$, which is in contrast with the single-frequency case ($c=0$), where multiple transitions are observed when we vary $\omega$.
﻿
\paragraph*{Fractional values of the ratio $\omega_2/\omega_1$}
﻿
We analyze the behavior of $\langle F \rangle$ for $c>1$, while $\omega=3$ is held fixed; we vary the parameter $c$ in small steps of $\Delta c=0.05$. Figure~\ref{fig:3} (a) shows the presence of multiple non-monotonic transitions at lower values of $c$, which gradually disappear when we increase $c$. Surprisingly, we find that within the range $2 < c <3$, there are several points at which $\langle F \rangle \approx 0$, while for $c=2$ and $c=3$ the time-averaged fidelity is non-zero. We observe a similar behavior for $3 \leq c \leq 4$. These observations indicate that for small fractional values of $c>1$, where the period $T > 2\pi/\omega$, the system exhibits characteristics that indicate an ergodic behavior. However, for $c \geq 7$, we observe $\langle F \rangle \neq 0$, which suggests the presence of scarring features that persist irrespective of whether $c$ is fractional or integer.

The density plot of \(\langle F \rangle\) in the $(\lambda_0,c)$-plane confirms the above observation, see Fig.~\ref{fig:3} (b). Scarring features are not significant at small \(\lambda_0\) for any \(c\). However, for \(\lambda_0 > 6\), they become increasingly prominent. Notably, as \(c\) increases, more rational values of \(c\) begin to exhibit scar-induced oscillations.

We can qualitatively understand the above behavior by analyzing the effective hamiltonian $H_F(1)$, wherein for $p \gg q$, the Bessel functions in $H^{(b)}_F$ drop to zero, $J_{-p\beta}(-\lambda_0/\omega)$,  $J_{q\beta}(-\lambda_0/c\omega)$ $\approx 0$, unless $\omega \ll 1$; this indicates that the dynamics is predominantly governed by $H^{(a)}_F$ (similar to the $c$ integer case). We remark that this behavior is markedly different from the case when $c$ is small. For the integer values of $c$, the parameter $p$ remains small, while $q=1$. However, when $c$ is a fraction, both $p$ and $q$ can be large, which highlights the significant role of Bessel functions in governing the behavior of $H_F(1)$ (and consequently the dynamics). Hence, the structure of $H_F(1)$ becomes highly dependent on $c$. 

﻿
﻿
\paragraph*{Approximation of a quasi-periodic drive}
﻿
Next to demonstrate the versatility of a two frequency drive with $c=p/q$ (a rational number), we use it to approximate a quasi-periodically driven Rydberg chain with $c\equiv\alpha=(1+\sqrt{5})/2$, a golden ratio. 
We refer to~\cite{dutta2025prethermalization} for a detailed numerical and analytical account of the dynamics under this protocol, in which a high-frequency Magnus expansion was implemented to obtain a renormalized PXP model that was then further used to illustrate a scar-induced prethermalization. However, the Magnus expansion is not convergent in the low-frequency regime. Also, the FPT can not be directly implemented to construct an effective Hamiltonian. We suggest that a better approach to address this issue is to approximate the quasi-periodic driving with a (two-frequency) periodic drive.
﻿

We achieve this through a rational approximation of $\alpha$ that involves the use of a Fibonacci chain (Pingala-Hemachandra sequence) and considering the successive ratios of its numbers. We discard the first few numbers and consider the following approximations to $\alpha$: $\alpha_1=8/5$, $\alpha_2=13/8$, $\alpha_3=21/13$, and $\alpha_4=34/21$ to compute the corresponding $\langle F \rangle$ as a function of $\omega$.
﻿

Figure~\ref{fig:4} (a) and (b) show that for $\alpha_1=8/5$, $\langle F \rangle$ is non-zero, indicating the presence of scar-induced oscillations in the frequency range $2 < \omega < 3.25$, but this disappears for the other values of $\alpha$ and the dynamics approaches to that of a quasi-periodic drive. Here, we observe that $c=\alpha_4=34/21$ is able to capture the true fidelity behavior reasonably well. In the high frequency regime, $\langle F \rangle$ exhibits identical behavior for all the four values of $\alpha$. It is possible to obtain the Floquet hamiltonian, while carefully accounting for the periodicity of the drive. The structure of $H_F$ is used to predict the key features of the actual dynamics~\cite{supplemental}. We remark that even for small fractional values of $c$, $H_F$ contains higher order Bessel functions, thereby distinguishing it from the case when $c$ is a small integer. Thus, the interplay of the drive period and the magnitude of $c$ gives a generic control on the scar-induced oscillations.
﻿
\paragraph*{Designing robust subharmonic responses}
﻿
\begin{figure}
	\includegraphics[width=0.95\linewidth]{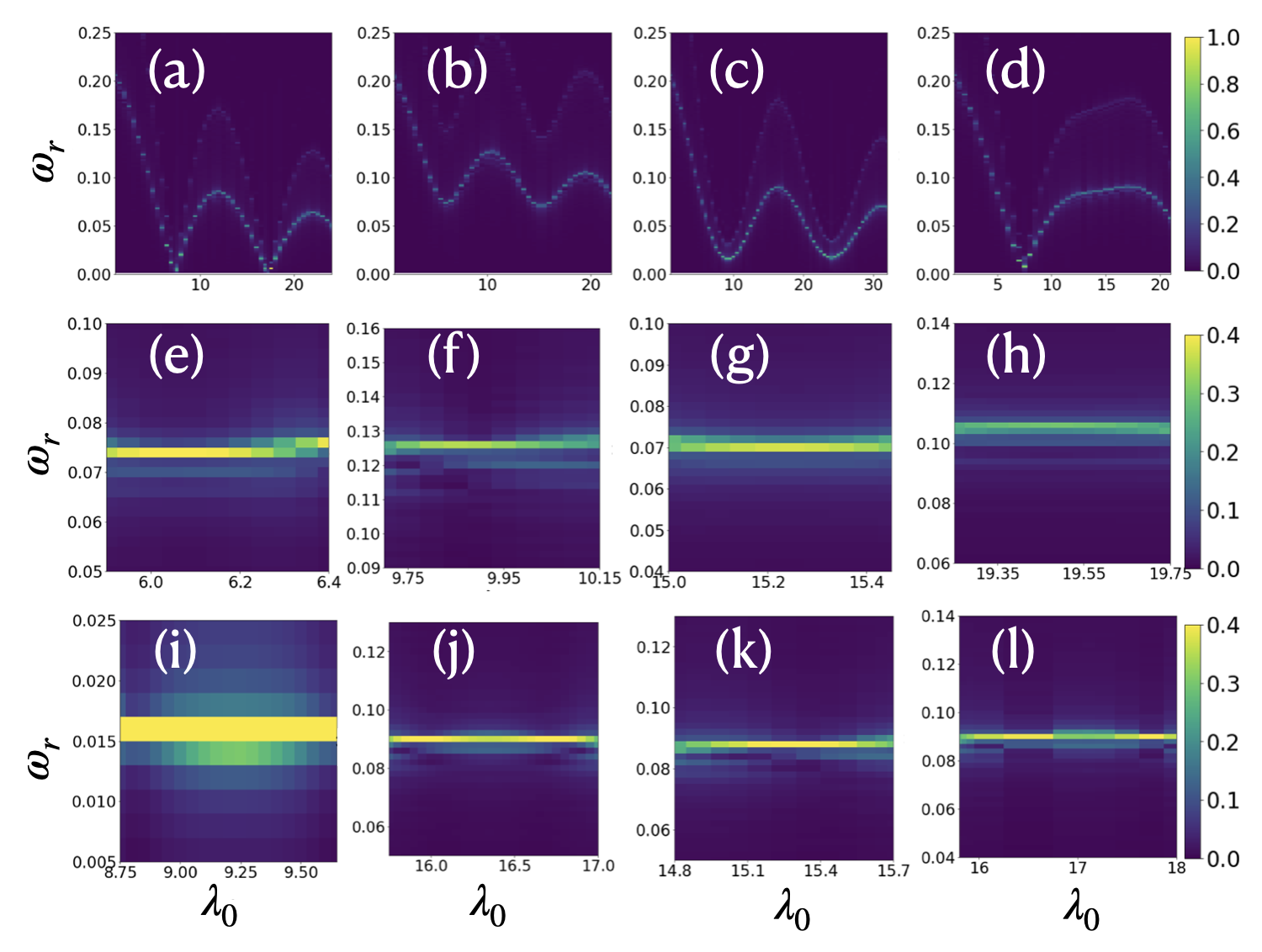}
	\caption{\textbf{Existence of subharmonics}. Density plot of the Fourier transform of magnetization exhibits an oscillatory variation of response frequency with $\lambda_0$ for $\omega = 2\pi$. Top panel: (a) $c=0$, (b) $c=2$, (c) $c=3$, and (d) $c=4$.  Middle panel (e)-(h), bottom panel (i)-(j) and (k)-(l) for $c=2$, $3$ and $4$, respectively, illustrate the stability of specific response frequencies and the presence of higher-order subharmonics.
	}
	\label{fig:subhar}
\end{figure}
﻿
We use the $z-$component of magnetization, $O(t)=\sigma^1_z(t)=\braket{\psi(t)|\sigma^1_z|\psi(t)}$, to detect the emergence of fractional and higher-order responses, if any, in a suitable high-frequency and high-amplitude regime. The presence of any subharmonic response serves as an indication of the spontaneous breaking of time-translation symmetry, with $\braket{O(t+nT)} = \braket{O(t)}$, where $n>1$ can be an integer or a fraction. Alternatively, we can regard this as long-time oscillations of the observable with frequency $\omega/n$.
﻿

﻿
In Fig.~\ref{fig:subhar} (a), (b), (c), and (d), we show the density plot of the absolute value of the Fourier transform of $O(t)$ for $c=0$, $2$, $3$ and $4$, respectively, while the drive frequency is kept fixed at $\omega=2\pi$ ($\nu_1=1$). Clearly, the response frequency, $\omega_r$, exhibits an oscillatory behavior and the intensity fluctuates, as we vary $\lambda_0$. For $c=2$, both the fluctuating intensity and $\omega_r$ do not drop to zero, which suggests the absence of thermalization at high drive amplitudes. We observe similar features for $c=3$ and $4$, where $\omega_r$ can come very close to zero, without exhibiting ergodicity. This is in contrast to the $c=0$ case for which $\omega_r$ drops to zero, the response is captured by the Bessel function $J_0(\lambda_0/\omega)$ that vanishes at $\lambda_0/\omega \approx 2.33, \, 5.48, \hdots$, indicating thermalization. 
However, for $c=2$, we observe peaks at $\lambda_0/\omega=2.33$ and $5.48$ that correspond to $\omega_r=\omega/12.5$, i.e., $\nu_r=2/25\nu$ (plot not shown).
﻿

The observed frequencies represent subharmonics, but the stability of the subharmoics is not uniform across the entire range of $\lambda_0$. For $c=2$, Fig.~\ref{fig:subhar} (e), (f), (g), and (h) show that the stable responses occur at $\omega_r/\omega \approx 3/40$, $1/8$, $1/7$, and $13/125$, respectively. For $c=3$, Fig.~\ref{fig:subhar} (i) and (j) show the stable responses at $\omega_r/\omega \approx 2/125$ and $\omega_r/\omega \approx 9/100$, respectively, whereas Fig.~\ref{fig:subhar} (k) and (l) show the stable responses for $c=4$ occurring at $\omega_r/\omega=11/125$ and $\omega_r/\omega=9/100$, respectively. Interestingly, we find that these are comparatively more stable than those observed for $c=2$. Also, the stability of a very high order response at $\omega_r/\omega = 2/125$ for $c=3$ is a remarkable feature, which is not so prominent for other values of $c$. Our results suggests that a two-frequency drive provides a better platform for realizing fractional time crystals.

\paragraph*{Conclusion} The two-frequency drive with ratio $c$ provides an effective control over the scar-lifetime and allows for the generation of complex dynamical features. We demonstrate that the scar induced oscillations survive even at low frequencies by tuning $c$ to small integers. However, the situation changes dramatically, when $c$ becomes a fraction or $c \gg 1$. Also, a rational value of $c$ allows us to capture the dynamics of a quasi-periodic drive. Furthermore, the higher order subharmonic response exhibited by the scar-induced oscillations is more stable for non-zero $c$, such a feature has not been explored in Floquet engineering. This control is potentially useful, as scar-dominated dynamics plays a crucial role in quantum state preparation. Many of these responses have the potential to be utilized in the quantum-enhanced sensing~\cite{moon2024discrete, yousefjani2025discrete, cabot2024continuous, montenegro2025quantum, iemini2024floquet}. Moreover, the two-frequency drive is expected to provide a rich physics in other scar models as well~\cite{mohapatra2023pronounced,yao2022quantum,schecter2019weak,moudgalya2018entanglement,lerose2025theory,sanada2023quantum}. Also, it would be interesting to study the fate of scars under multi-frequency drive protocols.

﻿
﻿
\subsection*{}
\subsection*{Acknowlegments} Authors acknowledge National Supercomputing Mission (NSM) for providing computing resources of ‘PARAM Shakti’ at IIT Kharagpur, which is implemented by C-DAC and supported by the Ministry of Electronics and Information Technology (MeitY) and Department of Science and Technology (DST).
﻿
﻿
﻿
﻿
\bibliographystyle{apsrev4-2}%
\bibliography{ref}

\begin{thebibliography}{86}%
\makeatletter
\providecommand \@ifxundefined [1]{%
 \@ifx{#1\undefined}
}%
\providecommand \@ifnum [1]{%
 \ifnum #1\expandafter \@firstoftwo
 \else \expandafter \@secondoftwo
 \fi
}%
\providecommand \@ifx [1]{%
 \ifx #1\expandafter \@firstoftwo
 \else \expandafter \@secondoftwo
 \fi
}%
\providecommand \natexlab [1]{#1}%
\providecommand \enquote  [1]{``#1''}%
\providecommand \bibnamefont  [1]{#1}%
\providecommand \bibfnamefont [1]{#1}%
\providecommand \citenamefont [1]{#1}%
\providecommand \href@noop [0]{\@secondoftwo}%
\providecommand \href [0]{\begingroup \@sanitize@url \@href}%
\providecommand \@href[1]{\@@startlink{#1}\@@href}%
\providecommand \@@href[1]{\endgroup#1\@@endlink}%
\providecommand \@sanitize@url [0]{\catcode `\\12\catcode `\$12\catcode
  `\&12\catcode `\#12\catcode `\^12\catcode `\_12\catcode `\%12\relax}%
\providecommand \@@startlink[1]{}%
\providecommand \@@endlink[0]{}%
\providecommand \url  [0]{\begingroup\@sanitize@url \@url }%
\providecommand \@url [1]{\endgroup\@href {#1}{\urlprefix }}%
\providecommand \urlprefix  [0]{URL }%
\providecommand \Eprint [0]{\href }%
\providecommand \doibase [0]{https://doi.org/}%
\providecommand \selectlanguage [0]{\@gobble}%
\providecommand \bibinfo  [0]{\@secondoftwo}%
\providecommand \bibfield  [0]{\@secondoftwo}%
\providecommand \translation [1]{[#1]}%
\providecommand \BibitemOpen [0]{}%
\providecommand \bibitemStop [0]{}%
\providecommand \bibitemNoStop [0]{.\EOS\space}%
\providecommand \EOS [0]{\spacefactor3000\relax}%
\providecommand \BibitemShut  [1]{\csname bibitem#1\endcsname}%
\let\auto@bib@innerbib\@empty
\bibitem [{\citenamefont {Lindner}\ \emph {et~al.}(2011)\citenamefont
  {Lindner}, \citenamefont {Refael},\ and\ \citenamefont
  {Galitski}}]{lindner2011floquet}%
  \BibitemOpen
  \bibfield  {author} {\bibinfo {author} {\bibfnamefont {N.~H.}\ \bibnamefont
  {Lindner}}, \bibinfo {author} {\bibfnamefont {G.}~\bibnamefont {Refael}},\
  and\ \bibinfo {author} {\bibfnamefont {V.}~\bibnamefont {Galitski}},\
  }\href@noop {} {\bibfield  {journal} {\bibinfo  {journal} {Nature Physics}\
  }\textbf {\bibinfo {volume} {7}},\ \bibinfo {pages} {490} (\bibinfo {year}
  {2011})}\BibitemShut {NoStop}%
\bibitem [{\citenamefont {Kundu}\ \emph {et~al.}(2014)\citenamefont {Kundu},
  \citenamefont {Fertig},\ and\ \citenamefont {Seradjeh}}]{kundu2014effective}%
  \BibitemOpen
  \bibfield  {author} {\bibinfo {author} {\bibfnamefont {A.}~\bibnamefont
  {Kundu}}, \bibinfo {author} {\bibfnamefont {H.}~\bibnamefont {Fertig}},\ and\
  \bibinfo {author} {\bibfnamefont {B.}~\bibnamefont {Seradjeh}},\ }\href@noop
  {} {\bibfield  {journal} {\bibinfo  {journal} {Physical review letters}\
  }\textbf {\bibinfo {volume} {113}},\ \bibinfo {pages} {236803} (\bibinfo
  {year} {2014})}\BibitemShut {NoStop}%
\bibitem [{\citenamefont {Else}\ \emph {et~al.}(2016)\citenamefont {Else},
  \citenamefont {Bauer},\ and\ \citenamefont {Nayak}}]{else2016floquet}%
  \BibitemOpen
  \bibfield  {author} {\bibinfo {author} {\bibfnamefont {D.~V.}\ \bibnamefont
  {Else}}, \bibinfo {author} {\bibfnamefont {B.}~\bibnamefont {Bauer}},\ and\
  \bibinfo {author} {\bibfnamefont {C.}~\bibnamefont {Nayak}},\ }\href@noop {}
  {\bibfield  {journal} {\bibinfo  {journal} {Physical review letters}\
  }\textbf {\bibinfo {volume} {117}},\ \bibinfo {pages} {090402} (\bibinfo
  {year} {2016})}\BibitemShut {NoStop}%
\bibitem [{\citenamefont {Aditya}\ and\ \citenamefont
  {Sen}(2023)}]{aditya2023dynamical}%
  \BibitemOpen
  \bibfield  {author} {\bibinfo {author} {\bibfnamefont {S.}~\bibnamefont
  {Aditya}}\ and\ \bibinfo {author} {\bibfnamefont {D.}~\bibnamefont {Sen}},\
  }\href@noop {} {\bibfield  {journal} {\bibinfo  {journal} {SciPost Physics
  Core}\ }\textbf {\bibinfo {volume} {6}},\ \bibinfo {pages} {083} (\bibinfo
  {year} {2023})}\BibitemShut {NoStop}%
\bibitem [{\citenamefont {Zhang}\ \emph {et~al.}(2017)\citenamefont {Zhang},
  \citenamefont {Hess}, \citenamefont {Kyprianidis}, \citenamefont {Becker},
  \citenamefont {Lee}, \citenamefont {Smith}, \citenamefont {Pagano},
  \citenamefont {Potirniche}, \citenamefont {Potter}, \citenamefont
  {Vishwanath} \emph {et~al.}}]{zhang2017observation}%
  \BibitemOpen
  \bibfield  {author} {\bibinfo {author} {\bibfnamefont {J.}~\bibnamefont
  {Zhang}}, \bibinfo {author} {\bibfnamefont {P.~W.}\ \bibnamefont {Hess}},
  \bibinfo {author} {\bibfnamefont {A.}~\bibnamefont {Kyprianidis}}, \bibinfo
  {author} {\bibfnamefont {P.}~\bibnamefont {Becker}}, \bibinfo {author}
  {\bibfnamefont {A.}~\bibnamefont {Lee}}, \bibinfo {author} {\bibfnamefont
  {J.}~\bibnamefont {Smith}}, \bibinfo {author} {\bibfnamefont
  {G.}~\bibnamefont {Pagano}}, \bibinfo {author} {\bibfnamefont {I.-D.}\
  \bibnamefont {Potirniche}}, \bibinfo {author} {\bibfnamefont {A.~C.}\
  \bibnamefont {Potter}}, \bibinfo {author} {\bibfnamefont {A.}~\bibnamefont
  {Vishwanath}}, \emph {et~al.},\ }\href@noop {} {\bibfield  {journal}
  {\bibinfo  {journal} {Nature}\ }\textbf {\bibinfo {volume} {543}},\ \bibinfo
  {pages} {217} (\bibinfo {year} {2017})}\BibitemShut {NoStop}%
\bibitem [{\citenamefont {Gangopadhay}\ and\ \citenamefont
  {Choudhury}(2025)}]{gangopadhay2025counterdiabatic}%
  \BibitemOpen
  \bibfield  {author} {\bibinfo {author} {\bibfnamefont {N.}~\bibnamefont
  {Gangopadhay}}\ and\ \bibinfo {author} {\bibfnamefont {S.}~\bibnamefont
  {Choudhury}},\ }\href@noop {} {\bibfield  {journal} {\bibinfo  {journal}
  {Physical Review Letters}\ }\textbf {\bibinfo {volume} {135}},\ \bibinfo
  {pages} {020407} (\bibinfo {year} {2025})}\BibitemShut {NoStop}%
\bibitem [{\citenamefont {Guo}\ \emph {et~al.}(2025)\citenamefont {Guo},
  \citenamefont {Mukherjee},\ and\ \citenamefont
  {Chowdhury}}]{guo2025dynamical}%
  \BibitemOpen
  \bibfield  {author} {\bibinfo {author} {\bibfnamefont {H.}~\bibnamefont
  {Guo}}, \bibinfo {author} {\bibfnamefont {R.}~\bibnamefont {Mukherjee}},\
  and\ \bibinfo {author} {\bibfnamefont {D.}~\bibnamefont {Chowdhury}},\
  }\href@noop {} {\bibfield  {journal} {\bibinfo  {journal} {Physical Review
  Letters}\ }\textbf {\bibinfo {volume} {134}},\ \bibinfo {pages} {226501}
  (\bibinfo {year} {2025})}\BibitemShut {NoStop}%
\bibitem [{\citenamefont {Oka}\ and\ \citenamefont
  {Kitamura}(2019)}]{oka2019floquet}%
  \BibitemOpen
  \bibfield  {author} {\bibinfo {author} {\bibfnamefont {T.}~\bibnamefont
  {Oka}}\ and\ \bibinfo {author} {\bibfnamefont {S.}~\bibnamefont {Kitamura}},\
  }\href@noop {} {\bibfield  {journal} {\bibinfo  {journal} {Annual Review of
  Condensed Matter Physics}\ }\textbf {\bibinfo {volume} {10}},\ \bibinfo
  {pages} {387} (\bibinfo {year} {2019})}\BibitemShut {NoStop}%
\bibitem [{\citenamefont {Bai}\ and\ \citenamefont
  {An}(2023)}]{bai2023floquet}%
  \BibitemOpen
  \bibfield  {author} {\bibinfo {author} {\bibfnamefont {S.-Y.}\ \bibnamefont
  {Bai}}\ and\ \bibinfo {author} {\bibfnamefont {J.-H.}\ \bibnamefont {An}},\
  }\href@noop {} {\bibfield  {journal} {\bibinfo  {journal} {Phys. Rev. Lett.}\
  }\textbf {\bibinfo {volume} {131}},\ \bibinfo {pages} {050801} (\bibinfo
  {year} {2023})}\BibitemShut {NoStop}%
\bibitem [{\citenamefont {Cayssol}\ \emph {et~al.}(2013)\citenamefont
  {Cayssol}, \citenamefont {D{\'o}ra}, \citenamefont {Simon},\ and\
  \citenamefont {Moessner}}]{cayssol2013floquet}%
  \BibitemOpen
  \bibfield  {author} {\bibinfo {author} {\bibfnamefont {J.}~\bibnamefont
  {Cayssol}}, \bibinfo {author} {\bibfnamefont {B.}~\bibnamefont {D{\'o}ra}},
  \bibinfo {author} {\bibfnamefont {F.}~\bibnamefont {Simon}},\ and\ \bibinfo
  {author} {\bibfnamefont {R.}~\bibnamefont {Moessner}},\ }\href@noop {}
  {\bibfield  {journal} {\bibinfo  {journal} {Phys. Status Solidi RRL}\
  }\textbf {\bibinfo {volume} {7}},\ \bibinfo {pages} {101} (\bibinfo {year}
  {2013})}\BibitemShut {NoStop}%
\bibitem [{\citenamefont {Rudner}\ and\ \citenamefont
  {Lindner}(2020)}]{rudner2020band}%
  \BibitemOpen
  \bibfield  {author} {\bibinfo {author} {\bibfnamefont {M.~S.}\ \bibnamefont
  {Rudner}}\ and\ \bibinfo {author} {\bibfnamefont {N.~H.}\ \bibnamefont
  {Lindner}},\ }\href@noop {} {\bibfield  {journal} {\bibinfo  {journal} {Nat.
  Rev. Phys}\ }\textbf {\bibinfo {volume} {2}},\ \bibinfo {pages} {229}
  (\bibinfo {year} {2020})}\BibitemShut {NoStop}%
\bibitem [{\citenamefont {Sacha}\ and\ \citenamefont
  {Zakrzewski}(2017)}]{sacha2017time}%
  \BibitemOpen
  \bibfield  {author} {\bibinfo {author} {\bibfnamefont {K.}~\bibnamefont
  {Sacha}}\ and\ \bibinfo {author} {\bibfnamefont {J.}~\bibnamefont
  {Zakrzewski}},\ }\href@noop {} {\bibfield  {journal} {\bibinfo  {journal}
  {Rep. Prog. Phys.}\ }\textbf {\bibinfo {volume} {81}},\ \bibinfo {pages}
  {016401} (\bibinfo {year} {2017})}\BibitemShut {NoStop}%
\bibitem [{\citenamefont {Srednicki}(1994)}]{srednicki1994chaos}%
  \BibitemOpen
  \bibfield  {author} {\bibinfo {author} {\bibfnamefont {M.}~\bibnamefont
  {Srednicki}},\ }\href@noop {} {\bibfield  {journal} {\bibinfo  {journal}
  {Phys. Rev. E}\ }\textbf {\bibinfo {volume} {50}},\ \bibinfo {pages} {888}
  (\bibinfo {year} {1994})}\BibitemShut {NoStop}%
\bibitem [{\citenamefont {D'Alessio}\ \emph {et~al.}(2016)\citenamefont
  {D'Alessio}, \citenamefont {Kafri}, \citenamefont {Polkovnikov},\ and\
  \citenamefont {Rigol}}]{d2016quantum}%
  \BibitemOpen
  \bibfield  {author} {\bibinfo {author} {\bibfnamefont {L.}~\bibnamefont
  {D'Alessio}}, \bibinfo {author} {\bibfnamefont {Y.}~\bibnamefont {Kafri}},
  \bibinfo {author} {\bibfnamefont {A.}~\bibnamefont {Polkovnikov}},\ and\
  \bibinfo {author} {\bibfnamefont {M.}~\bibnamefont {Rigol}},\ }\href@noop {}
  {\bibfield  {journal} {\bibinfo  {journal} {Adv. Phys.}\ }\textbf {\bibinfo
  {volume} {65}},\ \bibinfo {pages} {239} (\bibinfo {year} {2016})}\BibitemShut
  {NoStop}%
\bibitem [{\citenamefont {Mori}\ \emph {et~al.}(2018)\citenamefont {Mori},
  \citenamefont {Ikeda}, \citenamefont {Kaminishi},\ and\ \citenamefont
  {Ueda}}]{mori2018thermalization}%
  \BibitemOpen
  \bibfield  {author} {\bibinfo {author} {\bibfnamefont {T.}~\bibnamefont
  {Mori}}, \bibinfo {author} {\bibfnamefont {T.~N.}\ \bibnamefont {Ikeda}},
  \bibinfo {author} {\bibfnamefont {E.}~\bibnamefont {Kaminishi}},\ and\
  \bibinfo {author} {\bibfnamefont {M.}~\bibnamefont {Ueda}},\ }\href@noop {}
  {\bibfield  {journal} {\bibinfo  {journal} {J. Phys. B}\ }\textbf {\bibinfo
  {volume} {51}},\ \bibinfo {pages} {112001} (\bibinfo {year}
  {2018})}\BibitemShut {NoStop}%
\bibitem [{\citenamefont {Lazarides}\ \emph {et~al.}(2014)\citenamefont
  {Lazarides}, \citenamefont {Das},\ and\ \citenamefont
  {Moessner}}]{lazarides2014equilibrium}%
  \BibitemOpen
  \bibfield  {author} {\bibinfo {author} {\bibfnamefont {A.}~\bibnamefont
  {Lazarides}}, \bibinfo {author} {\bibfnamefont {A.}~\bibnamefont {Das}},\
  and\ \bibinfo {author} {\bibfnamefont {R.}~\bibnamefont {Moessner}},\
  }\href@noop {} {\bibfield  {journal} {\bibinfo  {journal} {Physical Review
  E}\ }\textbf {\bibinfo {volume} {90}},\ \bibinfo {pages} {012110} (\bibinfo
  {year} {2014})}\BibitemShut {NoStop}%
\bibitem [{\citenamefont {D’Alessio}\ and\ \citenamefont
  {Polkovnikov}(2013)}]{d2013many}%
  \BibitemOpen
  \bibfield  {author} {\bibinfo {author} {\bibfnamefont {L.}~\bibnamefont
  {D’Alessio}}\ and\ \bibinfo {author} {\bibfnamefont {A.}~\bibnamefont
  {Polkovnikov}},\ }\href@noop {} {\bibfield  {journal} {\bibinfo  {journal}
  {Annals of Physics}\ }\textbf {\bibinfo {volume} {333}},\ \bibinfo {pages}
  {19} (\bibinfo {year} {2013})}\BibitemShut {NoStop}%
\bibitem [{\citenamefont {D’Alessio}\ and\ \citenamefont
  {Rigol}(2014)}]{d2014long}%
  \BibitemOpen
  \bibfield  {author} {\bibinfo {author} {\bibfnamefont {L.}~\bibnamefont
  {D’Alessio}}\ and\ \bibinfo {author} {\bibfnamefont {M.}~\bibnamefont
  {Rigol}},\ }\href@noop {} {\bibfield  {journal} {\bibinfo  {journal}
  {Physical Review X}\ }\textbf {\bibinfo {volume} {4}},\ \bibinfo {pages}
  {041048} (\bibinfo {year} {2014})}\BibitemShut {NoStop}%
\bibitem [{\citenamefont {Ho}\ \emph {et~al.}(2023)\citenamefont {Ho},
  \citenamefont {Mori}, \citenamefont {Abanin},\ and\ \citenamefont
  {Dalla~Torre}}]{ho2023quantum}%
  \BibitemOpen
  \bibfield  {author} {\bibinfo {author} {\bibfnamefont {W.~W.}\ \bibnamefont
  {Ho}}, \bibinfo {author} {\bibfnamefont {T.}~\bibnamefont {Mori}}, \bibinfo
  {author} {\bibfnamefont {D.~A.}\ \bibnamefont {Abanin}},\ and\ \bibinfo
  {author} {\bibfnamefont {E.~G.}\ \bibnamefont {Dalla~Torre}},\ }\href@noop {}
  {\bibfield  {journal} {\bibinfo  {journal} {Annals of Physics}\ ,\ \bibinfo
  {pages} {169297}} (\bibinfo {year} {2023})}\BibitemShut {NoStop}%
\bibitem [{\citenamefont {Canovi}\ \emph {et~al.}(2016)\citenamefont {Canovi},
  \citenamefont {Kollar},\ and\ \citenamefont
  {Eckstein}}]{canovi2016stroboscopic}%
  \BibitemOpen
  \bibfield  {author} {\bibinfo {author} {\bibfnamefont {E.}~\bibnamefont
  {Canovi}}, \bibinfo {author} {\bibfnamefont {M.}~\bibnamefont {Kollar}},\
  and\ \bibinfo {author} {\bibfnamefont {M.}~\bibnamefont {Eckstein}},\
  }\href@noop {} {\bibfield  {journal} {\bibinfo  {journal} {Phys. Rev. E}\
  }\textbf {\bibinfo {volume} {93}},\ \bibinfo {pages} {012130} (\bibinfo
  {year} {2016})}\BibitemShut {NoStop}%
\bibitem [{\citenamefont {Abanin}\ \emph {et~al.}(2017)\citenamefont {Abanin},
  \citenamefont {De~Roeck}, \citenamefont {Ho},\ and\ \citenamefont
  {Huveneers}}]{abanin2017effective}%
  \BibitemOpen
  \bibfield  {author} {\bibinfo {author} {\bibfnamefont {D.~A.}\ \bibnamefont
  {Abanin}}, \bibinfo {author} {\bibfnamefont {W.}~\bibnamefont {De~Roeck}},
  \bibinfo {author} {\bibfnamefont {W.~W.}\ \bibnamefont {Ho}},\ and\ \bibinfo
  {author} {\bibfnamefont {F.}~\bibnamefont {Huveneers}},\ }\href@noop {}
  {\bibfield  {journal} {\bibinfo  {journal} {Phys. Rev. B}\ }\textbf {\bibinfo
  {volume} {95}},\ \bibinfo {pages} {014112} (\bibinfo {year}
  {2017})}\BibitemShut {NoStop}%
\bibitem [{\citenamefont {Bukov}\ \emph {et~al.}(2015)\citenamefont {Bukov},
  \citenamefont {D'Alessio},\ and\ \citenamefont
  {Polkovnikov}}]{bukov2015universal}%
  \BibitemOpen
  \bibfield  {author} {\bibinfo {author} {\bibfnamefont {M.}~\bibnamefont
  {Bukov}}, \bibinfo {author} {\bibfnamefont {L.}~\bibnamefont {D'Alessio}},\
  and\ \bibinfo {author} {\bibfnamefont {A.}~\bibnamefont {Polkovnikov}},\
  }\href@noop {} {\bibfield  {journal} {\bibinfo  {journal} {Adv. Phys.}\
  }\textbf {\bibinfo {volume} {64}},\ \bibinfo {pages} {139} (\bibinfo {year}
  {2015})}\BibitemShut {NoStop}%
\bibitem [{\citenamefont {Bukov}\ \emph {et~al.}(2016)\citenamefont {Bukov},
  \citenamefont {Heyl}, \citenamefont {Huse},\ and\ \citenamefont
  {Polkovnikov}}]{bukov2016heating}%
  \BibitemOpen
  \bibfield  {author} {\bibinfo {author} {\bibfnamefont {M.}~\bibnamefont
  {Bukov}}, \bibinfo {author} {\bibfnamefont {M.}~\bibnamefont {Heyl}},
  \bibinfo {author} {\bibfnamefont {D.~A.}\ \bibnamefont {Huse}},\ and\
  \bibinfo {author} {\bibfnamefont {A.}~\bibnamefont {Polkovnikov}},\
  }\href@noop {} {\bibfield  {journal} {\bibinfo  {journal} {Phys. Rev. B}\
  }\textbf {\bibinfo {volume} {93}},\ \bibinfo {pages} {155132} (\bibinfo
  {year} {2016})}\BibitemShut {NoStop}%
\bibitem [{\citenamefont {Abanin}\ \emph {et~al.}(2015)\citenamefont {Abanin},
  \citenamefont {De~Roeck},\ and\ \citenamefont
  {Huveneers}}]{abanin2015exponentially}%
  \BibitemOpen
  \bibfield  {author} {\bibinfo {author} {\bibfnamefont {D.~A.}\ \bibnamefont
  {Abanin}}, \bibinfo {author} {\bibfnamefont {W.}~\bibnamefont {De~Roeck}},\
  and\ \bibinfo {author} {\bibfnamefont {F.}~\bibnamefont {Huveneers}},\
  }\href@noop {} {\bibfield  {journal} {\bibinfo  {journal} {Phys. Rev. Lett.}\
  }\textbf {\bibinfo {volume} {115}},\ \bibinfo {pages} {256803} (\bibinfo
  {year} {2015})}\BibitemShut {NoStop}%
\bibitem [{\citenamefont {Ikeda}\ \emph {et~al.}(2022)\citenamefont {Ikeda},
  \citenamefont {Kitamura},\ and\ \citenamefont {Morimoto}}]{ikeda2022floquet}%
  \BibitemOpen
  \bibfield  {author} {\bibinfo {author} {\bibfnamefont {Y.}~\bibnamefont
  {Ikeda}}, \bibinfo {author} {\bibfnamefont {S.}~\bibnamefont {Kitamura}},\
  and\ \bibinfo {author} {\bibfnamefont {T.}~\bibnamefont {Morimoto}},\
  }\href@noop {} {\bibfield  {journal} {\bibinfo  {journal} {Progress of
  Theoretical and Experimental Physics}\ }\textbf {\bibinfo {volume} {2022}},\
  \bibinfo {pages} {04A101} (\bibinfo {year} {2022})}\BibitemShut {NoStop}%
\bibitem [{\citenamefont {Pe{\~n}a}\ \emph {et~al.}(2024)\citenamefont
  {Pe{\~n}a}, \citenamefont {Torres},\ and\ \citenamefont
  {Romero}}]{pena2024steering}%
  \BibitemOpen
  \bibfield  {author} {\bibinfo {author} {\bibfnamefont {R.}~\bibnamefont
  {Pe{\~n}a}}, \bibinfo {author} {\bibfnamefont {F.}~\bibnamefont {Torres}},\
  and\ \bibinfo {author} {\bibfnamefont {G.}~\bibnamefont {Romero}},\
  }\href@noop {} {\bibfield  {journal} {\bibinfo  {journal} {arXiv preprint
  arXiv:2401.03889}\ } (\bibinfo {year} {2024})}\BibitemShut {NoStop}%
\bibitem [{\citenamefont {Kar}(2017)}]{kar2017two}%
  \BibitemOpen
  \bibfield  {author} {\bibinfo {author} {\bibfnamefont {S.}~\bibnamefont
  {Kar}},\ }\href@noop {} {\bibfield  {journal} {\bibinfo  {journal} {Physical
  Review B}\ }\textbf {\bibinfo {volume} {95}},\ \bibinfo {pages} {085141}
  (\bibinfo {year} {2017})}\BibitemShut {NoStop}%
\bibitem [{\citenamefont {Kar}\ \emph {et~al.}(2016)\citenamefont {Kar},
  \citenamefont {Mukherjee},\ and\ \citenamefont {Sengupta}}]{kar2016tuning}%
  \BibitemOpen
  \bibfield  {author} {\bibinfo {author} {\bibfnamefont {S.}~\bibnamefont
  {Kar}}, \bibinfo {author} {\bibfnamefont {B.}~\bibnamefont {Mukherjee}},\
  and\ \bibinfo {author} {\bibfnamefont {K.}~\bibnamefont {Sengupta}},\
  }\href@noop {} {\bibfield  {journal} {\bibinfo  {journal} {Physical Review
  B}\ }\textbf {\bibinfo {volume} {94}},\ \bibinfo {pages} {075130} (\bibinfo
  {year} {2016})}\BibitemShut {NoStop}%
\bibitem [{\citenamefont {Else}\ \emph {et~al.}(2020)\citenamefont {Else},
  \citenamefont {Ho},\ and\ \citenamefont {Dumitrescu}}]{else2020long}%
  \BibitemOpen
  \bibfield  {author} {\bibinfo {author} {\bibfnamefont {D.~V.}\ \bibnamefont
  {Else}}, \bibinfo {author} {\bibfnamefont {W.~W.}\ \bibnamefont {Ho}},\ and\
  \bibinfo {author} {\bibfnamefont {P.~T.}\ \bibnamefont {Dumitrescu}},\
  }\href@noop {} {\bibfield  {journal} {\bibinfo  {journal} {Physical Review
  X}\ }\textbf {\bibinfo {volume} {10}},\ \bibinfo {pages} {021032} (\bibinfo
  {year} {2020})}\BibitemShut {NoStop}%
\bibitem [{\citenamefont {Das}\ \emph {et~al.}(2023)\citenamefont {Das},
  \citenamefont {Bhakuni}, \citenamefont {Santos},\ and\ \citenamefont
  {Sharma}}]{das2023periodically}%
  \BibitemOpen
  \bibfield  {author} {\bibinfo {author} {\bibfnamefont {P.}~\bibnamefont
  {Das}}, \bibinfo {author} {\bibfnamefont {D.~S.}\ \bibnamefont {Bhakuni}},
  \bibinfo {author} {\bibfnamefont {L.~F.}\ \bibnamefont {Santos}},\ and\
  \bibinfo {author} {\bibfnamefont {A.}~\bibnamefont {Sharma}},\ }\href@noop {}
  {\bibfield  {journal} {\bibinfo  {journal} {Physical Review A}\ }\textbf
  {\bibinfo {volume} {108}},\ \bibinfo {pages} {063716} (\bibinfo {year}
  {2023})}\BibitemShut {NoStop}%
\bibitem [{\citenamefont {Mori}\ \emph {et~al.}(2021)\citenamefont {Mori},
  \citenamefont {Zhao}, \citenamefont {Mintert}, \citenamefont {Knolle},\ and\
  \citenamefont {Moessner}}]{mori2021rigorous}%
  \BibitemOpen
  \bibfield  {author} {\bibinfo {author} {\bibfnamefont {T.}~\bibnamefont
  {Mori}}, \bibinfo {author} {\bibfnamefont {H.}~\bibnamefont {Zhao}}, \bibinfo
  {author} {\bibfnamefont {F.}~\bibnamefont {Mintert}}, \bibinfo {author}
  {\bibfnamefont {J.}~\bibnamefont {Knolle}},\ and\ \bibinfo {author}
  {\bibfnamefont {R.}~\bibnamefont {Moessner}},\ }\href@noop {} {\bibfield
  {journal} {\bibinfo  {journal} {Physical review letters}\ }\textbf {\bibinfo
  {volume} {127}},\ \bibinfo {pages} {050602} (\bibinfo {year}
  {2021})}\BibitemShut {NoStop}%
\bibitem [{\citenamefont {Kumar}\ and\ \citenamefont
  {Choudhury}(2024)}]{kumar2024prethermalization}%
  \BibitemOpen
  \bibfield  {author} {\bibinfo {author} {\bibfnamefont {S.}~\bibnamefont
  {Kumar}}\ and\ \bibinfo {author} {\bibfnamefont {S.}~\bibnamefont
  {Choudhury}},\ }\href@noop {} {\bibfield  {journal} {\bibinfo  {journal}
  {Physical Review E}\ }\textbf {\bibinfo {volume} {110}},\ \bibinfo {pages}
  {064150} (\bibinfo {year} {2024})}\BibitemShut {NoStop}%
\bibitem [{\citenamefont {Yan}\ \emph {et~al.}(2024)\citenamefont {Yan},
  \citenamefont {Moessner},\ and\ \citenamefont
  {Zhao}}]{yan2024prethermalization}%
  \BibitemOpen
  \bibfield  {author} {\bibinfo {author} {\bibfnamefont {J.}~\bibnamefont
  {Yan}}, \bibinfo {author} {\bibfnamefont {R.}~\bibnamefont {Moessner}},\ and\
  \bibinfo {author} {\bibfnamefont {H.}~\bibnamefont {Zhao}},\ }\href@noop {}
  {\bibfield  {journal} {\bibinfo  {journal} {Phys. Rev. B}\ }\textbf {\bibinfo
  {volume} {109}},\ \bibinfo {pages} {064305} (\bibinfo {year}
  {2024})}\BibitemShut {NoStop}%
\bibitem [{\citenamefont {Martin}\ \emph {et~al.}(2017)\citenamefont {Martin},
  \citenamefont {Refael},\ and\ \citenamefont
  {Halperin}}]{martin2017topological}%
  \BibitemOpen
  \bibfield  {author} {\bibinfo {author} {\bibfnamefont {I.}~\bibnamefont
  {Martin}}, \bibinfo {author} {\bibfnamefont {G.}~\bibnamefont {Refael}},\
  and\ \bibinfo {author} {\bibfnamefont {B.}~\bibnamefont {Halperin}},\
  }\href@noop {} {\bibfield  {journal} {\bibinfo  {journal} {Physical Review
  X}\ }\textbf {\bibinfo {volume} {7}},\ \bibinfo {pages} {041008} (\bibinfo
  {year} {2017})}\BibitemShut {NoStop}%
\bibitem [{\citenamefont {Qi}\ \emph {et~al.}(2024)\citenamefont {Qi},
  \citenamefont {Na}, \citenamefont {Refael},\ and\ \citenamefont
  {Peng}}]{qi2024real}%
  \BibitemOpen
  \bibfield  {author} {\bibinfo {author} {\bibfnamefont {Z.}~\bibnamefont
  {Qi}}, \bibinfo {author} {\bibfnamefont {I.}~\bibnamefont {Na}}, \bibinfo
  {author} {\bibfnamefont {G.}~\bibnamefont {Refael}},\ and\ \bibinfo {author}
  {\bibfnamefont {Y.}~\bibnamefont {Peng}},\ }\href@noop {} {\bibfield
  {journal} {\bibinfo  {journal} {Physical Review B}\ }\textbf {\bibinfo
  {volume} {110}},\ \bibinfo {pages} {014309} (\bibinfo {year}
  {2024})}\BibitemShut {NoStop}%
\bibitem [{\citenamefont {Long}\ \emph {et~al.}(2022)\citenamefont {Long},
  \citenamefont {Crowley},\ and\ \citenamefont {Chandran}}]{long2022many}%
  \BibitemOpen
  \bibfield  {author} {\bibinfo {author} {\bibfnamefont {D.~M.}\ \bibnamefont
  {Long}}, \bibinfo {author} {\bibfnamefont {P.~J.}\ \bibnamefont {Crowley}},\
  and\ \bibinfo {author} {\bibfnamefont {A.}~\bibnamefont {Chandran}},\
  }\href@noop {} {\bibfield  {journal} {\bibinfo  {journal} {Physical Review
  B}\ }\textbf {\bibinfo {volume} {105}},\ \bibinfo {pages} {144204} (\bibinfo
  {year} {2022})}\BibitemShut {NoStop}%
\bibitem [{\citenamefont {Long}\ \emph {et~al.}(2021)\citenamefont {Long},
  \citenamefont {Crowley},\ and\ \citenamefont
  {Chandran}}]{long2021nonadiabatic}%
  \BibitemOpen
  \bibfield  {author} {\bibinfo {author} {\bibfnamefont {D.~M.}\ \bibnamefont
  {Long}}, \bibinfo {author} {\bibfnamefont {P.~J.}\ \bibnamefont {Crowley}},\
  and\ \bibinfo {author} {\bibfnamefont {A.}~\bibnamefont {Chandran}},\
  }\href@noop {} {\bibfield  {journal} {\bibinfo  {journal} {Physical Review
  Letters}\ }\textbf {\bibinfo {volume} {126}},\ \bibinfo {pages} {106805}
  (\bibinfo {year} {2021})}\BibitemShut {NoStop}%
\bibitem [{\citenamefont {Crowley}\ \emph {et~al.}(2019)\citenamefont
  {Crowley}, \citenamefont {Martin},\ and\ \citenamefont
  {Chandran}}]{crowley2019topological}%
  \BibitemOpen
  \bibfield  {author} {\bibinfo {author} {\bibfnamefont {P.~J.}\ \bibnamefont
  {Crowley}}, \bibinfo {author} {\bibfnamefont {I.}~\bibnamefont {Martin}},\
  and\ \bibinfo {author} {\bibfnamefont {A.}~\bibnamefont {Chandran}},\
  }\href@noop {} {\bibfield  {journal} {\bibinfo  {journal} {Physical Review
  B}\ }\textbf {\bibinfo {volume} {99}},\ \bibinfo {pages} {064306} (\bibinfo
  {year} {2019})}\BibitemShut {NoStop}%
\bibitem [{\citenamefont {Lin}\ \emph {et~al.}(2016)\citenamefont {Lin},
  \citenamefont {Xiao}, \citenamefont {Yuan},\ and\ \citenamefont
  {Fan}}]{lin2016photonic}%
  \BibitemOpen
  \bibfield  {author} {\bibinfo {author} {\bibfnamefont {Q.}~\bibnamefont
  {Lin}}, \bibinfo {author} {\bibfnamefont {M.}~\bibnamefont {Xiao}}, \bibinfo
  {author} {\bibfnamefont {L.}~\bibnamefont {Yuan}},\ and\ \bibinfo {author}
  {\bibfnamefont {S.}~\bibnamefont {Fan}},\ }\href@noop {} {\bibfield
  {journal} {\bibinfo  {journal} {Nature communications}\ }\textbf {\bibinfo
  {volume} {7}},\ \bibinfo {pages} {13731} (\bibinfo {year}
  {2016})}\BibitemShut {NoStop}%
\bibitem [{\citenamefont {Ozawa}\ \emph {et~al.}(2016)\citenamefont {Ozawa},
  \citenamefont {Price}, \citenamefont {Goldman}, \citenamefont {Zilberberg},\
  and\ \citenamefont {Carusotto}}]{ozawa2016synthetic}%
  \BibitemOpen
  \bibfield  {author} {\bibinfo {author} {\bibfnamefont {T.}~\bibnamefont
  {Ozawa}}, \bibinfo {author} {\bibfnamefont {H.~M.}\ \bibnamefont {Price}},
  \bibinfo {author} {\bibfnamefont {N.}~\bibnamefont {Goldman}}, \bibinfo
  {author} {\bibfnamefont {O.}~\bibnamefont {Zilberberg}},\ and\ \bibinfo
  {author} {\bibfnamefont {I.}~\bibnamefont {Carusotto}},\ }\href@noop {}
  {\bibfield  {journal} {\bibinfo  {journal} {Physical Review A}\ }\textbf
  {\bibinfo {volume} {93}},\ \bibinfo {pages} {043827} (\bibinfo {year}
  {2016})}\BibitemShut {NoStop}%
\bibitem [{\citenamefont {Yuan}\ \emph {et~al.}(2016)\citenamefont {Yuan},
  \citenamefont {Shi},\ and\ \citenamefont {Fan}}]{yuan2016photonic}%
  \BibitemOpen
  \bibfield  {author} {\bibinfo {author} {\bibfnamefont {L.}~\bibnamefont
  {Yuan}}, \bibinfo {author} {\bibfnamefont {Y.}~\bibnamefont {Shi}},\ and\
  \bibinfo {author} {\bibfnamefont {S.}~\bibnamefont {Fan}},\ }\href@noop {}
  {\bibfield  {journal} {\bibinfo  {journal} {Optics letters}\ }\textbf
  {\bibinfo {volume} {41}},\ \bibinfo {pages} {741} (\bibinfo {year}
  {2016})}\BibitemShut {NoStop}%
\bibitem [{\citenamefont {Murakami}\ \emph {et~al.}(2023)\citenamefont
  {Murakami}, \citenamefont {Sch{\"u}ler}, \citenamefont {Arita},\ and\
  \citenamefont {Werner}}]{murakami2023suppression}%
  \BibitemOpen
  \bibfield  {author} {\bibinfo {author} {\bibfnamefont {Y.}~\bibnamefont
  {Murakami}}, \bibinfo {author} {\bibfnamefont {M.}~\bibnamefont
  {Sch{\"u}ler}}, \bibinfo {author} {\bibfnamefont {R.}~\bibnamefont {Arita}},\
  and\ \bibinfo {author} {\bibfnamefont {P.}~\bibnamefont {Werner}},\
  }\href@noop {} {\bibfield  {journal} {\bibinfo  {journal} {Physical Review
  B}\ }\textbf {\bibinfo {volume} {108}},\ \bibinfo {pages} {035151} (\bibinfo
  {year} {2023})}\BibitemShut {NoStop}%
\bibitem [{\citenamefont {Chen}\ \emph {et~al.}(2024)\citenamefont {Chen},
  \citenamefont {Zhu},\ and\ \citenamefont {Viebahn}}]{chen2024mitigating}%
  \BibitemOpen
  \bibfield  {author} {\bibinfo {author} {\bibfnamefont {Y.}~\bibnamefont
  {Chen}}, \bibinfo {author} {\bibfnamefont {Z.}~\bibnamefont {Zhu}},\ and\
  \bibinfo {author} {\bibfnamefont {K.}~\bibnamefont {Viebahn}},\ }\href@noop
  {} {\bibfield  {journal} {\bibinfo  {journal} {arXiv preprint
  arXiv:2410.12308}\ } (\bibinfo {year} {2024})}\BibitemShut {NoStop}%
\bibitem [{\citenamefont {Brise{\~n}o-Colunga}\ \emph
  {et~al.}(2025)\citenamefont {Brise{\~n}o-Colunga}, \citenamefont {Bhandari},
  \citenamefont {Das}, \citenamefont {Nguyen}, \citenamefont {Kim},
  \citenamefont {Santiago}, \citenamefont {Siddiqi}, \citenamefont {Jordan},\
  and\ \citenamefont {Dressel}}]{briseno2025dynamical}%
  \BibitemOpen
  \bibfield  {author} {\bibinfo {author} {\bibfnamefont {D.~D.}\ \bibnamefont
  {Brise{\~n}o-Colunga}}, \bibinfo {author} {\bibfnamefont {B.}~\bibnamefont
  {Bhandari}}, \bibinfo {author} {\bibfnamefont {D.}~\bibnamefont {Das}},
  \bibinfo {author} {\bibfnamefont {L.~B.}\ \bibnamefont {Nguyen}}, \bibinfo
  {author} {\bibfnamefont {Y.}~\bibnamefont {Kim}}, \bibinfo {author}
  {\bibfnamefont {D.~I.}\ \bibnamefont {Santiago}}, \bibinfo {author}
  {\bibfnamefont {I.}~\bibnamefont {Siddiqi}}, \bibinfo {author} {\bibfnamefont
  {A.~N.}\ \bibnamefont {Jordan}},\ and\ \bibinfo {author} {\bibfnamefont
  {J.}~\bibnamefont {Dressel}},\ }\href@noop {} {\bibfield  {journal} {\bibinfo
   {journal} {arXiv preprint arXiv:2505.22606}\ } (\bibinfo {year}
  {2025})}\BibitemShut {NoStop}%
\bibitem [{\citenamefont {Hajati}\ and\ \citenamefont
  {Burkard}(2024)}]{hajati2024dynamic}%
  \BibitemOpen
  \bibfield  {author} {\bibinfo {author} {\bibfnamefont {Y.}~\bibnamefont
  {Hajati}}\ and\ \bibinfo {author} {\bibfnamefont {G.}~\bibnamefont
  {Burkard}},\ }\href@noop {} {\bibfield  {journal} {\bibinfo  {journal}
  {Physical Review B}\ }\textbf {\bibinfo {volume} {110}},\ \bibinfo {pages}
  {245301} (\bibinfo {year} {2024})}\BibitemShut {NoStop}%
\bibitem [{\citenamefont {Nandkishore}\ and\ \citenamefont
  {Huse}(2015)}]{nandkishore2015many}%
  \BibitemOpen
  \bibfield  {author} {\bibinfo {author} {\bibfnamefont {R.}~\bibnamefont
  {Nandkishore}}\ and\ \bibinfo {author} {\bibfnamefont {D.~A.}\ \bibnamefont
  {Huse}},\ }\href@noop {} {\bibfield  {journal} {\bibinfo  {journal} {Annu.
  Rev. Condens. Matter Phys.}\ }\textbf {\bibinfo {volume} {6}},\ \bibinfo
  {pages} {15} (\bibinfo {year} {2015})}\BibitemShut {NoStop}%
\bibitem [{\citenamefont {Alet}\ and\ \citenamefont
  {Laflorencie}(2018)}]{alet2018many}%
  \BibitemOpen
  \bibfield  {author} {\bibinfo {author} {\bibfnamefont {F.}~\bibnamefont
  {Alet}}\ and\ \bibinfo {author} {\bibfnamefont {N.}~\bibnamefont
  {Laflorencie}},\ }\href@noop {} {\bibfield  {journal} {\bibinfo  {journal}
  {Comptes Rendus Physique}\ }\textbf {\bibinfo {volume} {19}},\ \bibinfo
  {pages} {498} (\bibinfo {year} {2018})}\BibitemShut {NoStop}%
\bibitem [{\citenamefont {Altman}(2018)}]{altman2018many}%
  \BibitemOpen
  \bibfield  {author} {\bibinfo {author} {\bibfnamefont {E.}~\bibnamefont
  {Altman}},\ }\href@noop {} {\bibfield  {journal} {\bibinfo  {journal} {Nat.
  Phys.}\ }\textbf {\bibinfo {volume} {14}},\ \bibinfo {pages} {979} (\bibinfo
  {year} {2018})}\BibitemShut {NoStop}%
\bibitem [{\citenamefont {Lazarides}\ \emph {et~al.}(2015)\citenamefont
  {Lazarides}, \citenamefont {Das},\ and\ \citenamefont
  {Moessner}}]{lazarides2015fate}%
  \BibitemOpen
  \bibfield  {author} {\bibinfo {author} {\bibfnamefont {A.}~\bibnamefont
  {Lazarides}}, \bibinfo {author} {\bibfnamefont {A.}~\bibnamefont {Das}},\
  and\ \bibinfo {author} {\bibfnamefont {R.}~\bibnamefont {Moessner}},\
  }\href@noop {} {\bibfield  {journal} {\bibinfo  {journal} {Phys. Rev. Lett.}\
  }\textbf {\bibinfo {volume} {115}},\ \bibinfo {pages} {030402} (\bibinfo
  {year} {2015})}\BibitemShut {NoStop}%
\bibitem [{\citenamefont {Abanin}\ \emph {et~al.}(2019)\citenamefont {Abanin},
  \citenamefont {Altman}, \citenamefont {Bloch},\ and\ \citenamefont
  {Serbyn}}]{abanin2019colloquium}%
  \BibitemOpen
  \bibfield  {author} {\bibinfo {author} {\bibfnamefont {D.~A.}\ \bibnamefont
  {Abanin}}, \bibinfo {author} {\bibfnamefont {E.}~\bibnamefont {Altman}},
  \bibinfo {author} {\bibfnamefont {I.}~\bibnamefont {Bloch}},\ and\ \bibinfo
  {author} {\bibfnamefont {M.}~\bibnamefont {Serbyn}},\ }\href@noop {}
  {\bibfield  {journal} {\bibinfo  {journal} {Rev. Mod. Phys.}\ }\textbf
  {\bibinfo {volume} {91}},\ \bibinfo {pages} {021001} (\bibinfo {year}
  {2019})}\BibitemShut {NoStop}%
\bibitem [{\citenamefont {Serbyn}\ \emph {et~al.}(2021)\citenamefont {Serbyn},
  \citenamefont {Abanin},\ and\ \citenamefont {Papi{\'c}}}]{serbyn2021quantum}%
  \BibitemOpen
  \bibfield  {author} {\bibinfo {author} {\bibfnamefont {M.}~\bibnamefont
  {Serbyn}}, \bibinfo {author} {\bibfnamefont {D.~A.}\ \bibnamefont {Abanin}},\
  and\ \bibinfo {author} {\bibfnamefont {Z.}~\bibnamefont {Papi{\'c}}},\
  }\href@noop {} {\bibfield  {journal} {\bibinfo  {journal} {Nature Physics}\
  }\textbf {\bibinfo {volume} {17}},\ \bibinfo {pages} {675} (\bibinfo {year}
  {2021})}\BibitemShut {NoStop}%
\bibitem [{\citenamefont {Turner}\ \emph {et~al.}(2018)\citenamefont {Turner},
  \citenamefont {Michailidis}, \citenamefont {Abanin}, \citenamefont {Serbyn},\
  and\ \citenamefont {Papi{\'c}}}]{turner2018weak}%
  \BibitemOpen
  \bibfield  {author} {\bibinfo {author} {\bibfnamefont {C.~J.}\ \bibnamefont
  {Turner}}, \bibinfo {author} {\bibfnamefont {A.~A.}\ \bibnamefont
  {Michailidis}}, \bibinfo {author} {\bibfnamefont {D.~A.}\ \bibnamefont
  {Abanin}}, \bibinfo {author} {\bibfnamefont {M.}~\bibnamefont {Serbyn}},\
  and\ \bibinfo {author} {\bibfnamefont {Z.}~\bibnamefont {Papi{\'c}}},\
  }\href@noop {} {\bibfield  {journal} {\bibinfo  {journal} {Nature Physics}\
  }\textbf {\bibinfo {volume} {14}},\ \bibinfo {pages} {745} (\bibinfo {year}
  {2018})}\BibitemShut {NoStop}%
\bibitem [{\citenamefont {Bernien}\ \emph {et~al.}(2017)\citenamefont
  {Bernien}, \citenamefont {Schwartz}, \citenamefont {Keesling}, \citenamefont
  {Levine}, \citenamefont {Omran}, \citenamefont {Pichler}, \citenamefont
  {Choi}, \citenamefont {Zibrov}, \citenamefont {Endres}, \citenamefont
  {Greiner} \emph {et~al.}}]{bernien2017probing}%
  \BibitemOpen
  \bibfield  {author} {\bibinfo {author} {\bibfnamefont {H.}~\bibnamefont
  {Bernien}}, \bibinfo {author} {\bibfnamefont {S.}~\bibnamefont {Schwartz}},
  \bibinfo {author} {\bibfnamefont {A.}~\bibnamefont {Keesling}}, \bibinfo
  {author} {\bibfnamefont {H.}~\bibnamefont {Levine}}, \bibinfo {author}
  {\bibfnamefont {A.}~\bibnamefont {Omran}}, \bibinfo {author} {\bibfnamefont
  {H.}~\bibnamefont {Pichler}}, \bibinfo {author} {\bibfnamefont
  {S.}~\bibnamefont {Choi}}, \bibinfo {author} {\bibfnamefont {A.~S.}\
  \bibnamefont {Zibrov}}, \bibinfo {author} {\bibfnamefont {M.}~\bibnamefont
  {Endres}}, \bibinfo {author} {\bibfnamefont {M.}~\bibnamefont {Greiner}},
  \emph {et~al.},\ }\href@noop {} {\bibfield  {journal} {\bibinfo  {journal}
  {Nature}\ }\textbf {\bibinfo {volume} {551}},\ \bibinfo {pages} {579}
  (\bibinfo {year} {2017})}\BibitemShut {NoStop}%
\bibitem [{\citenamefont {McClarty}\ \emph {et~al.}(2020)\citenamefont
  {McClarty}, \citenamefont {Haque}, \citenamefont {Sen},\ and\ \citenamefont
  {Richter}}]{mcclarty2020disorder}%
  \BibitemOpen
  \bibfield  {author} {\bibinfo {author} {\bibfnamefont {P.~A.}\ \bibnamefont
  {McClarty}}, \bibinfo {author} {\bibfnamefont {M.}~\bibnamefont {Haque}},
  \bibinfo {author} {\bibfnamefont {A.}~\bibnamefont {Sen}},\ and\ \bibinfo
  {author} {\bibfnamefont {J.}~\bibnamefont {Richter}},\ }\href@noop {}
  {\bibfield  {journal} {\bibinfo  {journal} {Phys. Rev. B}\ }\textbf {\bibinfo
  {volume} {102}},\ \bibinfo {pages} {224303} (\bibinfo {year}
  {2020})}\BibitemShut {NoStop}%
\bibitem [{\citenamefont {Mohapatra}\ and\ \citenamefont
  {Balram}(2023)}]{mohapatra2023pronounced}%
  \BibitemOpen
  \bibfield  {author} {\bibinfo {author} {\bibfnamefont {S.}~\bibnamefont
  {Mohapatra}}\ and\ \bibinfo {author} {\bibfnamefont {A.~C.}\ \bibnamefont
  {Balram}},\ }\href@noop {} {\bibfield  {journal} {\bibinfo  {journal} {Phys.
  Rev. B}\ }\textbf {\bibinfo {volume} {107}},\ \bibinfo {pages} {235121}
  (\bibinfo {year} {2023})}\BibitemShut {NoStop}%
\bibitem [{\citenamefont {Chandran}\ \emph {et~al.}(2023)\citenamefont
  {Chandran}, \citenamefont {Iadecola}, \citenamefont {Khemani},\ and\
  \citenamefont {Moessner}}]{chandran2023quantum}%
  \BibitemOpen
  \bibfield  {author} {\bibinfo {author} {\bibfnamefont {A.}~\bibnamefont
  {Chandran}}, \bibinfo {author} {\bibfnamefont {T.}~\bibnamefont {Iadecola}},
  \bibinfo {author} {\bibfnamefont {V.}~\bibnamefont {Khemani}},\ and\ \bibinfo
  {author} {\bibfnamefont {R.}~\bibnamefont {Moessner}},\ }\href@noop {}
  {\bibfield  {journal} {\bibinfo  {journal} {Annual Review of Condensed Matter
  Physics}\ }\textbf {\bibinfo {volume} {14}},\ \bibinfo {pages} {443}
  (\bibinfo {year} {2023})}\BibitemShut {NoStop}%
\bibitem [{\citenamefont {Zhao}\ \emph {et~al.}(2020)\citenamefont {Zhao},
  \citenamefont {Vovrosh}, \citenamefont {Mintert},\ and\ \citenamefont
  {Knolle}}]{zhao2020quantum}%
  \BibitemOpen
  \bibfield  {author} {\bibinfo {author} {\bibfnamefont {H.}~\bibnamefont
  {Zhao}}, \bibinfo {author} {\bibfnamefont {J.}~\bibnamefont {Vovrosh}},
  \bibinfo {author} {\bibfnamefont {F.}~\bibnamefont {Mintert}},\ and\ \bibinfo
  {author} {\bibfnamefont {J.}~\bibnamefont {Knolle}},\ }\href@noop {}
  {\bibfield  {journal} {\bibinfo  {journal} {Physical review letters}\
  }\textbf {\bibinfo {volume} {124}},\ \bibinfo {pages} {160604} (\bibinfo
  {year} {2020})}\BibitemShut {NoStop}%
\bibitem [{\citenamefont {Moudgalya}\ \emph {et~al.}(2022)\citenamefont
  {Moudgalya}, \citenamefont {Bernevig},\ and\ \citenamefont
  {Regnault}}]{moudgalya2022quantum}%
  \BibitemOpen
  \bibfield  {author} {\bibinfo {author} {\bibfnamefont {S.}~\bibnamefont
  {Moudgalya}}, \bibinfo {author} {\bibfnamefont {B.~A.}\ \bibnamefont
  {Bernevig}},\ and\ \bibinfo {author} {\bibfnamefont {N.}~\bibnamefont
  {Regnault}},\ }\href@noop {} {\bibfield  {journal} {\bibinfo  {journal} {Rep.
  Prog. Phys.}\ }\textbf {\bibinfo {volume} {85}},\ \bibinfo {pages} {086501}
  (\bibinfo {year} {2022})}\BibitemShut {NoStop}%
\bibitem [{\citenamefont {Moudgalya}\ and\ \citenamefont
  {Motrunich}(2022)}]{moudgalya2022hilbert}%
  \BibitemOpen
  \bibfield  {author} {\bibinfo {author} {\bibfnamefont {S.}~\bibnamefont
  {Moudgalya}}\ and\ \bibinfo {author} {\bibfnamefont {O.~I.}\ \bibnamefont
  {Motrunich}},\ }\href@noop {} {\bibfield  {journal} {\bibinfo  {journal}
  {Physical Review X}\ }\textbf {\bibinfo {volume} {12}},\ \bibinfo {pages}
  {011050} (\bibinfo {year} {2022})}\BibitemShut {NoStop}%
\bibitem [{\citenamefont {Choi}\ \emph {et~al.}(2019)\citenamefont {Choi},
  \citenamefont {Turner}, \citenamefont {Pichler}, \citenamefont {Ho},
  \citenamefont {Michailidis}, \citenamefont {Papi{\'c}}, \citenamefont
  {Serbyn}, \citenamefont {Lukin},\ and\ \citenamefont
  {Abanin}}]{choi2019emergent}%
  \BibitemOpen
  \bibfield  {author} {\bibinfo {author} {\bibfnamefont {S.}~\bibnamefont
  {Choi}}, \bibinfo {author} {\bibfnamefont {C.~J.}\ \bibnamefont {Turner}},
  \bibinfo {author} {\bibfnamefont {H.}~\bibnamefont {Pichler}}, \bibinfo
  {author} {\bibfnamefont {W.~W.}\ \bibnamefont {Ho}}, \bibinfo {author}
  {\bibfnamefont {A.~A.}\ \bibnamefont {Michailidis}}, \bibinfo {author}
  {\bibfnamefont {Z.}~\bibnamefont {Papi{\'c}}}, \bibinfo {author}
  {\bibfnamefont {M.}~\bibnamefont {Serbyn}}, \bibinfo {author} {\bibfnamefont
  {M.~D.}\ \bibnamefont {Lukin}},\ and\ \bibinfo {author} {\bibfnamefont
  {D.~A.}\ \bibnamefont {Abanin}},\ }\href@noop {} {\bibfield  {journal}
  {\bibinfo  {journal} {Phys. Rev. Lett.}\ }\textbf {\bibinfo {volume} {122}},\
  \bibinfo {pages} {220603} (\bibinfo {year} {2019})}\BibitemShut {NoStop}%
\bibitem [{\citenamefont {Zhang}\ \emph {et~al.}(2023)\citenamefont {Zhang},
  \citenamefont {Dong}, \citenamefont {Gao}, \citenamefont {Zhao},
  \citenamefont {Hao}, \citenamefont {Desaules}, \citenamefont {Guo},
  \citenamefont {Chen}, \citenamefont {Deng}, \citenamefont {Liu} \emph
  {et~al.}}]{zhang2023many}%
  \BibitemOpen
  \bibfield  {author} {\bibinfo {author} {\bibfnamefont {P.}~\bibnamefont
  {Zhang}}, \bibinfo {author} {\bibfnamefont {H.}~\bibnamefont {Dong}},
  \bibinfo {author} {\bibfnamefont {Y.}~\bibnamefont {Gao}}, \bibinfo {author}
  {\bibfnamefont {L.}~\bibnamefont {Zhao}}, \bibinfo {author} {\bibfnamefont
  {J.}~\bibnamefont {Hao}}, \bibinfo {author} {\bibfnamefont {J.-Y.}\
  \bibnamefont {Desaules}}, \bibinfo {author} {\bibfnamefont {Q.}~\bibnamefont
  {Guo}}, \bibinfo {author} {\bibfnamefont {J.}~\bibnamefont {Chen}}, \bibinfo
  {author} {\bibfnamefont {J.}~\bibnamefont {Deng}}, \bibinfo {author}
  {\bibfnamefont {B.}~\bibnamefont {Liu}}, \emph {et~al.},\ }\href@noop {}
  {\bibfield  {journal} {\bibinfo  {journal} {Nature Physics}\ }\textbf
  {\bibinfo {volume} {19}},\ \bibinfo {pages} {120} (\bibinfo {year}
  {2023})}\BibitemShut {NoStop}%
\bibitem [{\citenamefont {Dooley}(2021)}]{dooley2021robust}%
  \BibitemOpen
  \bibfield  {author} {\bibinfo {author} {\bibfnamefont {S.}~\bibnamefont
  {Dooley}},\ }\href@noop {} {\bibfield  {journal} {\bibinfo  {journal} {PRX
  Quantum}\ }\textbf {\bibinfo {volume} {2}},\ \bibinfo {pages} {020330}
  (\bibinfo {year} {2021})}\BibitemShut {NoStop}%
\bibitem [{\citenamefont {Dooley}\ \emph {et~al.}(2023)\citenamefont {Dooley},
  \citenamefont {Pappalardi},\ and\ \citenamefont
  {Goold}}]{dooley2023entanglement}%
  \BibitemOpen
  \bibfield  {author} {\bibinfo {author} {\bibfnamefont {S.}~\bibnamefont
  {Dooley}}, \bibinfo {author} {\bibfnamefont {S.}~\bibnamefont {Pappalardi}},\
  and\ \bibinfo {author} {\bibfnamefont {J.}~\bibnamefont {Goold}},\
  }\href@noop {} {\bibfield  {journal} {\bibinfo  {journal} {Physical Review
  B}\ }\textbf {\bibinfo {volume} {107}},\ \bibinfo {pages} {035123} (\bibinfo
  {year} {2023})}\BibitemShut {NoStop}%
\bibitem [{\citenamefont {Desaules}\ \emph {et~al.}(2022)\citenamefont
  {Desaules}, \citenamefont {Pietracaprina}, \citenamefont {Papi{\'c}},
  \citenamefont {Goold},\ and\ \citenamefont
  {Pappalardi}}]{desaules2022extensive}%
  \BibitemOpen
  \bibfield  {author} {\bibinfo {author} {\bibfnamefont {J.-Y.}\ \bibnamefont
  {Desaules}}, \bibinfo {author} {\bibfnamefont {F.}~\bibnamefont
  {Pietracaprina}}, \bibinfo {author} {\bibfnamefont {Z.}~\bibnamefont
  {Papi{\'c}}}, \bibinfo {author} {\bibfnamefont {J.}~\bibnamefont {Goold}},\
  and\ \bibinfo {author} {\bibfnamefont {S.}~\bibnamefont {Pappalardi}},\
  }\href@noop {} {\bibfield  {journal} {\bibinfo  {journal} {Physical Review
  Letters}\ }\textbf {\bibinfo {volume} {129}},\ \bibinfo {pages} {020601}
  (\bibinfo {year} {2022})}\BibitemShut {NoStop}%
\bibitem [{\citenamefont {Hudomal}\ \emph {et~al.}(2022)\citenamefont
  {Hudomal}, \citenamefont {Desaules}, \citenamefont {Mukherjee}, \citenamefont
  {Su}, \citenamefont {Halimeh},\ and\ \citenamefont
  {Papi{\'c}}}]{hudomal2022driving}%
  \BibitemOpen
  \bibfield  {author} {\bibinfo {author} {\bibfnamefont {A.}~\bibnamefont
  {Hudomal}}, \bibinfo {author} {\bibfnamefont {J.-Y.}\ \bibnamefont
  {Desaules}}, \bibinfo {author} {\bibfnamefont {B.}~\bibnamefont {Mukherjee}},
  \bibinfo {author} {\bibfnamefont {G.-X.}\ \bibnamefont {Su}}, \bibinfo
  {author} {\bibfnamefont {J.~C.}\ \bibnamefont {Halimeh}},\ and\ \bibinfo
  {author} {\bibfnamefont {Z.}~\bibnamefont {Papi{\'c}}},\ }\href@noop {}
  {\bibfield  {journal} {\bibinfo  {journal} {Phys. Rev. B}\ }\textbf {\bibinfo
  {volume} {106}},\ \bibinfo {pages} {104302} (\bibinfo {year}
  {2022})}\BibitemShut {NoStop}%
\bibitem [{\citenamefont {Maskara}\ \emph {et~al.}(2021)\citenamefont
  {Maskara}, \citenamefont {Michailidis}, \citenamefont {Ho}, \citenamefont
  {Bluvstein}, \citenamefont {Choi}, \citenamefont {Lukin},\ and\ \citenamefont
  {Serbyn}}]{maskara2021discrete}%
  \BibitemOpen
  \bibfield  {author} {\bibinfo {author} {\bibfnamefont {N.}~\bibnamefont
  {Maskara}}, \bibinfo {author} {\bibfnamefont {A.~A.}\ \bibnamefont
  {Michailidis}}, \bibinfo {author} {\bibfnamefont {W.~W.}\ \bibnamefont {Ho}},
  \bibinfo {author} {\bibfnamefont {D.}~\bibnamefont {Bluvstein}}, \bibinfo
  {author} {\bibfnamefont {S.}~\bibnamefont {Choi}}, \bibinfo {author}
  {\bibfnamefont {M.~D.}\ \bibnamefont {Lukin}},\ and\ \bibinfo {author}
  {\bibfnamefont {M.}~\bibnamefont {Serbyn}},\ }\href@noop {} {\bibfield
  {journal} {\bibinfo  {journal} {Physical Review Letters}\ }\textbf {\bibinfo
  {volume} {127}},\ \bibinfo {pages} {090602} (\bibinfo {year}
  {2021})}\BibitemShut {NoStop}%
\bibitem [{\citenamefont {Bluvstein}\ \emph {et~al.}(2021)\citenamefont
  {Bluvstein}, \citenamefont {Omran}, \citenamefont {Levine}, \citenamefont
  {Keesling}, \citenamefont {Semeghini}, \citenamefont {Ebadi}, \citenamefont
  {Wang}, \citenamefont {Michailidis}, \citenamefont {Maskara}, \citenamefont
  {Ho} \emph {et~al.}}]{bluvstein2021controlling}%
  \BibitemOpen
  \bibfield  {author} {\bibinfo {author} {\bibfnamefont {D.}~\bibnamefont
  {Bluvstein}}, \bibinfo {author} {\bibfnamefont {A.}~\bibnamefont {Omran}},
  \bibinfo {author} {\bibfnamefont {H.}~\bibnamefont {Levine}}, \bibinfo
  {author} {\bibfnamefont {A.}~\bibnamefont {Keesling}}, \bibinfo {author}
  {\bibfnamefont {G.}~\bibnamefont {Semeghini}}, \bibinfo {author}
  {\bibfnamefont {S.}~\bibnamefont {Ebadi}}, \bibinfo {author} {\bibfnamefont
  {T.~T.}\ \bibnamefont {Wang}}, \bibinfo {author} {\bibfnamefont {A.~A.}\
  \bibnamefont {Michailidis}}, \bibinfo {author} {\bibfnamefont
  {N.}~\bibnamefont {Maskara}}, \bibinfo {author} {\bibfnamefont {W.~W.}\
  \bibnamefont {Ho}}, \emph {et~al.},\ }\href@noop {} {\bibfield  {journal}
  {\bibinfo  {journal} {Science}\ }\textbf {\bibinfo {volume} {371}},\ \bibinfo
  {pages} {1355} (\bibinfo {year} {2021})}\BibitemShut {NoStop}%
\bibitem [{\citenamefont {Mukherjee}\ \emph {et~al.}(2020)\citenamefont
  {Mukherjee}, \citenamefont {Nandy}, \citenamefont {Sen}, \citenamefont
  {Sen},\ and\ \citenamefont {Sengupta}}]{mukherjee2020collapse}%
  \BibitemOpen
  \bibfield  {author} {\bibinfo {author} {\bibfnamefont {B.}~\bibnamefont
  {Mukherjee}}, \bibinfo {author} {\bibfnamefont {S.}~\bibnamefont {Nandy}},
  \bibinfo {author} {\bibfnamefont {A.}~\bibnamefont {Sen}}, \bibinfo {author}
  {\bibfnamefont {D.}~\bibnamefont {Sen}},\ and\ \bibinfo {author}
  {\bibfnamefont {K.}~\bibnamefont {Sengupta}},\ }\href@noop {} {\bibfield
  {journal} {\bibinfo  {journal} {Phys. Rev. B}\ }\textbf {\bibinfo {volume}
  {101}},\ \bibinfo {pages} {245107} (\bibinfo {year} {2020})}\BibitemShut
  {NoStop}%
\bibitem [{\citenamefont {Mukherjee}\ \emph {et~al.}(2022)\citenamefont
  {Mukherjee}, \citenamefont {Sen},\ and\ \citenamefont
  {Sengupta}}]{mukherjee2022periodically}%
  \BibitemOpen
  \bibfield  {author} {\bibinfo {author} {\bibfnamefont {B.}~\bibnamefont
  {Mukherjee}}, \bibinfo {author} {\bibfnamefont {A.}~\bibnamefont {Sen}},\
  and\ \bibinfo {author} {\bibfnamefont {K.}~\bibnamefont {Sengupta}},\
  }\href@noop {} {\bibfield  {journal} {\bibinfo  {journal} {Phys. Rev. B}\
  }\textbf {\bibinfo {volume} {106}},\ \bibinfo {pages} {064305} (\bibinfo
  {year} {2022})}\BibitemShut {NoStop}%
\bibitem [{\citenamefont {Huang}\ and\ \citenamefont
  {Li}(2024)}]{huang2024engineering}%
  \BibitemOpen
  \bibfield  {author} {\bibinfo {author} {\bibfnamefont {K.}~\bibnamefont
  {Huang}}\ and\ \bibinfo {author} {\bibfnamefont {X.}~\bibnamefont {Li}},\
  }\href@noop {} {\bibfield  {journal} {\bibinfo  {journal} {Physical Review
  B}\ }\textbf {\bibinfo {volume} {109}},\ \bibinfo {pages} {064306} (\bibinfo
  {year} {2024})}\BibitemShut {NoStop}%
\bibitem [{\citenamefont {Dutta}\ \emph {et~al.}(2025)\citenamefont {Dutta},
  \citenamefont {Choudhury},\ and\ \citenamefont
  {Shukla}}]{dutta2025prethermalization}%
  \BibitemOpen
  \bibfield  {author} {\bibinfo {author} {\bibfnamefont {P.}~\bibnamefont
  {Dutta}}, \bibinfo {author} {\bibfnamefont {S.}~\bibnamefont {Choudhury}},\
  and\ \bibinfo {author} {\bibfnamefont {V.}~\bibnamefont {Shukla}},\
  }\href@noop {} {\bibfield  {journal} {\bibinfo  {journal} {Physical Review
  B}\ }\textbf {\bibinfo {volume} {111}},\ \bibinfo {pages} {064303} (\bibinfo
  {year} {2025})}\BibitemShut {NoStop}%
\bibitem [{\citenamefont {Banerjee}\ \emph {et~al.}(2024)\citenamefont
  {Banerjee}, \citenamefont {Choudhury},\ and\ \citenamefont
  {Sengupta}}]{banerjee2024exact}%
  \BibitemOpen
  \bibfield  {author} {\bibinfo {author} {\bibfnamefont {T.}~\bibnamefont
  {Banerjee}}, \bibinfo {author} {\bibfnamefont {S.}~\bibnamefont
  {Choudhury}},\ and\ \bibinfo {author} {\bibfnamefont {K.}~\bibnamefont
  {Sengupta}},\ }\href@noop {} {\bibfield  {journal} {\bibinfo  {journal}
  {arXiv preprint arXiv:2404.06536}\ } (\bibinfo {year} {2024})}\BibitemShut
  {NoStop}%
\bibitem [{\citenamefont {Ghosh}\ \emph {et~al.}(2025)\citenamefont {Ghosh},
  \citenamefont {Choudhury}, \citenamefont {Sen},\ and\ \citenamefont
  {Sengupta}}]{ghosh2025heating}%
  \BibitemOpen
  \bibfield  {author} {\bibinfo {author} {\bibfnamefont {K.}~\bibnamefont
  {Ghosh}}, \bibinfo {author} {\bibfnamefont {S.}~\bibnamefont {Choudhury}},
  \bibinfo {author} {\bibfnamefont {D.}~\bibnamefont {Sen}},\ and\ \bibinfo
  {author} {\bibfnamefont {K.}~\bibnamefont {Sengupta}},\ }\href@noop {}
  {\bibfield  {journal} {\bibinfo  {journal} {arXiv preprint arXiv:2508.02783}\
  } (\bibinfo {year} {2025})}\BibitemShut {NoStop}%
\bibitem [{sup()}]{supplemental}%
  \BibitemOpen
  \href@noop {} {}\bibinfo {note} {See Supplemental Material.}\BibitemShut
  {Stop}%
\bibitem [{\citenamefont {Moon}\ \emph {et~al.}(2024)\citenamefont {Moon},
  \citenamefont {Schindler}, \citenamefont {Smith}, \citenamefont {Druga},
  \citenamefont {Zhang}, \citenamefont {Bukov},\ and\ \citenamefont
  {Ajoy}}]{moon2024discrete}%
  \BibitemOpen
  \bibfield  {author} {\bibinfo {author} {\bibfnamefont {L.~J.~I.}\
  \bibnamefont {Moon}}, \bibinfo {author} {\bibfnamefont {P.~M.}\ \bibnamefont
  {Schindler}}, \bibinfo {author} {\bibfnamefont {R.~J.}\ \bibnamefont
  {Smith}}, \bibinfo {author} {\bibfnamefont {E.}~\bibnamefont {Druga}},
  \bibinfo {author} {\bibfnamefont {Z.-R.}\ \bibnamefont {Zhang}}, \bibinfo
  {author} {\bibfnamefont {M.}~\bibnamefont {Bukov}},\ and\ \bibinfo {author}
  {\bibfnamefont {A.}~\bibnamefont {Ajoy}},\ }\href@noop {} {\bibfield
  {journal} {\bibinfo  {journal} {arXiv preprint arXiv:2410.05625}\ } (\bibinfo
  {year} {2024})}\BibitemShut {NoStop}%
\bibitem [{\citenamefont {Yousefjani}\ \emph {et~al.}(2025)\citenamefont
  {Yousefjani}, \citenamefont {Sacha},\ and\ \citenamefont
  {Bayat}}]{yousefjani2025discrete}%
  \BibitemOpen
  \bibfield  {author} {\bibinfo {author} {\bibfnamefont {R.}~\bibnamefont
  {Yousefjani}}, \bibinfo {author} {\bibfnamefont {K.}~\bibnamefont {Sacha}},\
  and\ \bibinfo {author} {\bibfnamefont {A.}~\bibnamefont {Bayat}},\
  }\href@noop {} {\bibfield  {journal} {\bibinfo  {journal} {Physical Review
  B}\ }\textbf {\bibinfo {volume} {111}},\ \bibinfo {pages} {125159} (\bibinfo
  {year} {2025})}\BibitemShut {NoStop}%
\bibitem [{\citenamefont {Cabot}\ \emph {et~al.}(2024)\citenamefont {Cabot},
  \citenamefont {Carollo},\ and\ \citenamefont
  {Lesanovsky}}]{cabot2024continuous}%
  \BibitemOpen
  \bibfield  {author} {\bibinfo {author} {\bibfnamefont {A.}~\bibnamefont
  {Cabot}}, \bibinfo {author} {\bibfnamefont {F.}~\bibnamefont {Carollo}},\
  and\ \bibinfo {author} {\bibfnamefont {I.}~\bibnamefont {Lesanovsky}},\
  }\href@noop {} {\bibfield  {journal} {\bibinfo  {journal} {Physical Review
  Letters}\ }\textbf {\bibinfo {volume} {132}},\ \bibinfo {pages} {050801}
  (\bibinfo {year} {2024})}\BibitemShut {NoStop}%
\bibitem [{\citenamefont {Montenegro}\ \emph {et~al.}(2025)\citenamefont
  {Montenegro}, \citenamefont {Mukhopadhyay}, \citenamefont {Yousefjani},
  \citenamefont {Sarkar}, \citenamefont {Mishra}, \citenamefont {Paris},\ and\
  \citenamefont {Bayat}}]{montenegro2025quantum}%
  \BibitemOpen
  \bibfield  {author} {\bibinfo {author} {\bibfnamefont {V.}~\bibnamefont
  {Montenegro}}, \bibinfo {author} {\bibfnamefont {C.}~\bibnamefont
  {Mukhopadhyay}}, \bibinfo {author} {\bibfnamefont {R.}~\bibnamefont
  {Yousefjani}}, \bibinfo {author} {\bibfnamefont {S.}~\bibnamefont {Sarkar}},
  \bibinfo {author} {\bibfnamefont {U.}~\bibnamefont {Mishra}}, \bibinfo
  {author} {\bibfnamefont {M.~G.}\ \bibnamefont {Paris}},\ and\ \bibinfo
  {author} {\bibfnamefont {A.}~\bibnamefont {Bayat}},\ }\href@noop {}
  {\bibfield  {journal} {\bibinfo  {journal} {Physics Reports}\ }\textbf
  {\bibinfo {volume} {1134}},\ \bibinfo {pages} {1} (\bibinfo {year}
  {2025})}\BibitemShut {NoStop}%
\bibitem [{\citenamefont {Iemini}\ \emph {et~al.}(2024)\citenamefont {Iemini},
  \citenamefont {Fazio},\ and\ \citenamefont {Sanpera}}]{iemini2024floquet}%
  \BibitemOpen
  \bibfield  {author} {\bibinfo {author} {\bibfnamefont {F.}~\bibnamefont
  {Iemini}}, \bibinfo {author} {\bibfnamefont {R.}~\bibnamefont {Fazio}},\ and\
  \bibinfo {author} {\bibfnamefont {A.}~\bibnamefont {Sanpera}},\ }\href@noop
  {} {\bibfield  {journal} {\bibinfo  {journal} {Physical Review A}\ }\textbf
  {\bibinfo {volume} {109}},\ \bibinfo {pages} {L050203} (\bibinfo {year}
  {2024})}\BibitemShut {NoStop}%
\bibitem [{\citenamefont {Yao}\ \emph {et~al.}(2022)\citenamefont {Yao},
  \citenamefont {Pan}, \citenamefont {Liu},\ and\ \citenamefont
  {Zhai}}]{yao2022quantum}%
  \BibitemOpen
  \bibfield  {author} {\bibinfo {author} {\bibfnamefont {Z.}~\bibnamefont
  {Yao}}, \bibinfo {author} {\bibfnamefont {L.}~\bibnamefont {Pan}}, \bibinfo
  {author} {\bibfnamefont {S.}~\bibnamefont {Liu}},\ and\ \bibinfo {author}
  {\bibfnamefont {H.}~\bibnamefont {Zhai}},\ }\href@noop {} {\bibfield
  {journal} {\bibinfo  {journal} {Physical Review B}\ }\textbf {\bibinfo
  {volume} {105}},\ \bibinfo {pages} {125123} (\bibinfo {year}
  {2022})}\BibitemShut {NoStop}%
\bibitem [{\citenamefont {Schecter}\ and\ \citenamefont
  {Iadecola}(2019)}]{schecter2019weak}%
  \BibitemOpen
  \bibfield  {author} {\bibinfo {author} {\bibfnamefont {M.}~\bibnamefont
  {Schecter}}\ and\ \bibinfo {author} {\bibfnamefont {T.}~\bibnamefont
  {Iadecola}},\ }\href@noop {} {\bibfield  {journal} {\bibinfo  {journal}
  {Physical review letters}\ }\textbf {\bibinfo {volume} {123}},\ \bibinfo
  {pages} {147201} (\bibinfo {year} {2019})}\BibitemShut {NoStop}%
\bibitem [{\citenamefont {Moudgalya}\ \emph {et~al.}(2018)\citenamefont
  {Moudgalya}, \citenamefont {Regnault},\ and\ \citenamefont
  {Bernevig}}]{moudgalya2018entanglement}%
  \BibitemOpen
  \bibfield  {author} {\bibinfo {author} {\bibfnamefont {S.}~\bibnamefont
  {Moudgalya}}, \bibinfo {author} {\bibfnamefont {N.}~\bibnamefont
  {Regnault}},\ and\ \bibinfo {author} {\bibfnamefont {B.~A.}\ \bibnamefont
  {Bernevig}},\ }\href@noop {} {\bibfield  {journal} {\bibinfo  {journal}
  {Physical Review B}\ }\textbf {\bibinfo {volume} {98}},\ \bibinfo {pages}
  {235156} (\bibinfo {year} {2018})}\BibitemShut {NoStop}%
\bibitem [{\citenamefont {Lerose}\ \emph {et~al.}(2025)\citenamefont {Lerose},
  \citenamefont {Parolini}, \citenamefont {Fazio}, \citenamefont {Abanin},\
  and\ \citenamefont {Pappalardi}}]{lerose2025theory}%
  \BibitemOpen
  \bibfield  {author} {\bibinfo {author} {\bibfnamefont {A.}~\bibnamefont
  {Lerose}}, \bibinfo {author} {\bibfnamefont {T.}~\bibnamefont {Parolini}},
  \bibinfo {author} {\bibfnamefont {R.}~\bibnamefont {Fazio}}, \bibinfo
  {author} {\bibfnamefont {D.~A.}\ \bibnamefont {Abanin}},\ and\ \bibinfo
  {author} {\bibfnamefont {S.}~\bibnamefont {Pappalardi}},\ }\href@noop {}
  {\bibfield  {journal} {\bibinfo  {journal} {Physical Review X}\ }\textbf
  {\bibinfo {volume} {15}},\ \bibinfo {pages} {011020} (\bibinfo {year}
  {2025})}\BibitemShut {NoStop}%
\bibitem [{\citenamefont {Sanada}\ \emph {et~al.}(2023)\citenamefont {Sanada},
  \citenamefont {Miao},\ and\ \citenamefont {Katsura}}]{sanada2023quantum}%
  \BibitemOpen
  \bibfield  {author} {\bibinfo {author} {\bibfnamefont {K.}~\bibnamefont
  {Sanada}}, \bibinfo {author} {\bibfnamefont {Y.}~\bibnamefont {Miao}},\ and\
  \bibinfo {author} {\bibfnamefont {H.}~\bibnamefont {Katsura}},\ }\href@noop
  {} {\bibfield  {journal} {\bibinfo  {journal} {Physical Review B}\ }\textbf
  {\bibinfo {volume} {108}},\ \bibinfo {pages} {155102} (\bibinfo {year}
  {2023})}\BibitemShut {NoStop}%
\bibitem [{\citenamefont {Sen}\ \emph {et~al.}(2021)\citenamefont {Sen},
  \citenamefont {Sen},\ and\ \citenamefont {Sengupta}}]{sen2021analytic}%
  \BibitemOpen
  \bibfield  {author} {\bibinfo {author} {\bibfnamefont {A.}~\bibnamefont
  {Sen}}, \bibinfo {author} {\bibfnamefont {D.}~\bibnamefont {Sen}},\ and\
  \bibinfo {author} {\bibfnamefont {K.}~\bibnamefont {Sengupta}},\ }\href@noop
  {} {\bibfield  {journal} {\bibinfo  {journal} {J. Phys. Condens. Matter}\
  }\textbf {\bibinfo {volume} {33}},\ \bibinfo {pages} {443003} (\bibinfo
  {year} {2021})}\BibitemShut {NoStop}%
\bibitem [{\citenamefont {Haldar}\ \emph {et~al.}(2021)\citenamefont {Haldar},
  \citenamefont {Sen}, \citenamefont {Moessner},\ and\ \citenamefont
  {Das}}]{haldar2021dynamical}%
  \BibitemOpen
  \bibfield  {author} {\bibinfo {author} {\bibfnamefont {A.}~\bibnamefont
  {Haldar}}, \bibinfo {author} {\bibfnamefont {D.}~\bibnamefont {Sen}},
  \bibinfo {author} {\bibfnamefont {R.}~\bibnamefont {Moessner}},\ and\
  \bibinfo {author} {\bibfnamefont {A.}~\bibnamefont {Das}},\ }\href@noop {}
  {\bibfield  {journal} {\bibinfo  {journal} {Phys. Rev. X}\ }\textbf {\bibinfo
  {volume} {11}},\ \bibinfo {pages} {021008} (\bibinfo {year}
  {2021})}\BibitemShut {NoStop}%
\end{thebibliography}%

\pagebreak

\onecolumngrid

\section*{Supplemental Material}

\section*{A. Effective Hamiltonian $H_F$ }
We provide a brief outline of the essential steps for the derivation of effective Hamiltonian. To start with, we use the eigenstates $\ket{n}$ of $S^z = \sum_i \sigma^z_{i}$ as a basis to write $H_0(t) \ket{n}= E_n(t) \ket{n}$, where $\braket{m|n}=\delta_{mn}$ and $E_n (t) = \bra{n} H_0(t) \ket{n}$. Therefore, the Schr\"odinger equation for the potential $V=0$
\begin{equation}\label{eq:SE1}
	i\hbar \frac{\partial}{\partial t}\ket{n(t)} = H(t) \ket{n(t)},
\end{equation}
admits exact solutions given by
\begin{equation} \label{eq:SE1_exactsol}
	\ket{n(t)} = e^{-i\int^t_0 E_n(t') dt'}\ket{n(0)}.
\end{equation}
Following Refs.~\cite{mukherjee2020collapse,sen2021analytic}, to first order in $V$, we obtain 
\begin{equation}
	\label{eq:degpert_HF}
	\braket{m|H_F(1)|n} = \frac{\braket{m|V|n}}{T}\int^T_0 dt \, e^{i \int^t_0 dt' \, [E_m(t') - E_n(t')]} .
\end{equation}

Next, we consider two states $\ket{m}$ and $\ket{n}$ so that $\braket{m|V|n} \neq 0$. Also, $V\ket{m} \sim \ket{m+1}+\ket{m-1}$; therefore, $E_m(t)-E_n(t)=\pm \lambda(t)$. Moreover, the use of the expression $\sigma^x = \tilde{\sigma}^+ + \tilde{\sigma}^-$, where $\tilde{\sigma}^{\pm}\ket{n}=\ket{n\pm 1}$, allows us to obtain $\braket{m|V|n}=-\Omega$.

\subsection*{Integral values of the ratio $c=\omega_2/\omega_1$}

\begin{figure}[h!]
	\centering
	\includegraphics[width=0.95\linewidth]{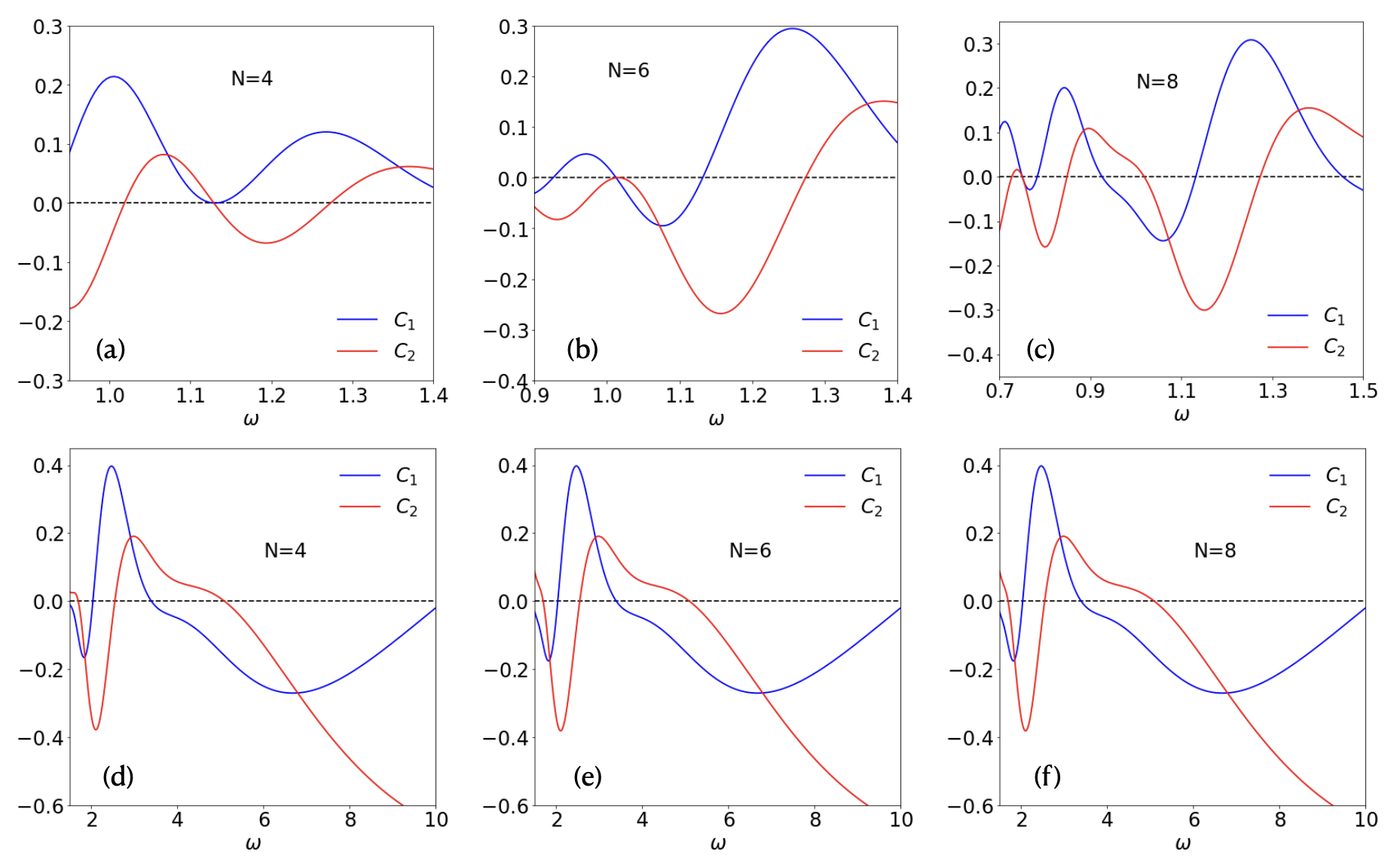}
	\caption{\textbf{Characteristics of $c = 3$}. Variation of $C_1$ and $C_2$ with $\omega$ for different $N$. \textit{Top panel:} Low-frequency regime — (a) $N = 4$, (b) $N = 6$, (c) $N = 8$. \textit{Bottom panel:} High-frequency regime — (d) $N = 4$, (e) $N = 6$, (f) $N = 8$. The roots of $C_1 = 0$ and $C_2 = 0$ correspond to the frequencies where ergodicity emerges. For $\omega \geq 1.4$, no such roots exist, indicating the absence of thermalization, and both $C_1$ and $C_2$ remain unaffected by $N$. In contrast, for $\omega < 1.4$, a single critical $\omega$ exists, which decreases slightly with increasing $N$.}
	\label{fig:convergec3}
\end{figure}

\begin{figure}[h!]
	\includegraphics[width=0.95\linewidth]{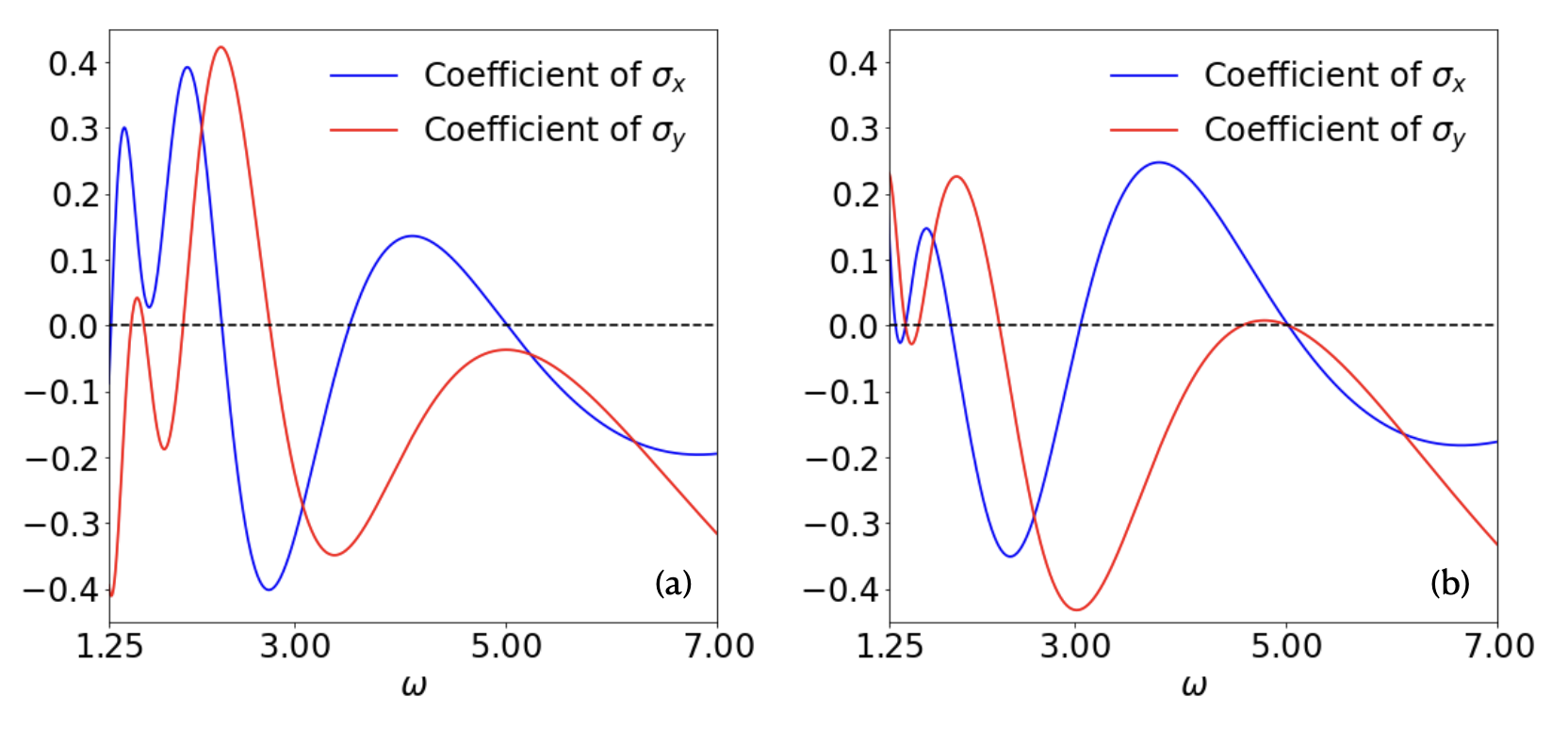}
	\caption{Coefficients of $\Tilde{\sigma}_{j}^{x}$ and $\Tilde{\sigma}_{j}^{y}$ for (a) $c = 4$ and (b) $c = 5$. For $c = 4$, the coefficients do not intersect along the dashed horizontal line, indicating the absence of ergodicity. For $c = 5$, two such intersections occur at $\omega \approx 1.28$ and $\omega \approx 5$. These results are in good agreement with numerical predictions. In both plots, $N = 8$.}
	\label{fig:convergec4and5N8}
\end{figure}

We use Eq.~\eqref{eq:degpert_HF} to obtain 
\begin{equation}
	(H_{F})_{mn} = -\frac{\Omega}{T} \int_{0}^{T}dt e^{i a \lambda_0\int_{0}^{t}[sin(\omega t^{\prime})+sin(c \omega t^{\prime})]dt^{\prime}},
\end{equation}
where $T=2\pi/\omega$ is the period of the drive with $\omega_1=\omega$ and $a=\pm 1$. We simplify the above integral by using the identity
\begin{equation}\label{eqn: Bessel}
	e^{ib_1(cos\omega t_1)+ib_2(cos\omega t_2)}  = \sum_{\alpha,\beta=-\infty}^{\infty} J_{\alpha}(b_1) J_{\beta}(b_2) i^{\alpha+\beta} e^{i(\alpha \omega_1+ \beta \omega_2)t}.
\end{equation}
This allows us to express the Floquet Hamiltonian as
\begin{equation}
	H_F(1) = \sum_{j}\sum_{n} \Bigl[\ket{n+1} (H_{F})_{n+1,n} \bra{n} + \ket{n-1} (H_{F})_{n-1,n} \bra{n} \Bigr],
\end{equation}
where 
\begin{equation}\nonumber
	(H_{F})_{n\pm 1} = -\frac{\Omega}{T}\int_{0}^{T} dt \sum_{\alpha,\beta} i^{\alpha+\beta} J_{\alpha}(\mp \lambda_{0}/\omega) J_{\beta}(\mp \lambda_{0}/c\omega) e^{\pm i\frac{\lambda_0}{\omega}(1+1/c)}  e^{i\omega(\alpha+\beta c)t}.
\end{equation}
We remark that the integral is non-zero for $\alpha+\beta c = 0$. After further simplification, the first order effective Hamiltonian is given by
\begin{equation}\label{eq:HF1}
	H_{F}(1) = H_{F}^{(a)}+H_{F}^{(b)}    
\end{equation}
where
\begin{equation}\nonumber
	H_{F}^{(a)}  = -J_{0}\left(\frac{\lambda_0}{\omega}\right) J_{0}\left(\frac{\lambda_0}{c \omega}\right)  
	\biggl[ \cos\left( \frac{(c+1)\lambda_0}{c\omega} \right) \sum_{j} \tilde{\sigma}_{j}^{x} 
	  - \sin\left( \frac{(c+1)\lambda_0}{c\omega} \right) \sum_{j} \tilde{\sigma}_{j}^{y} \biggr]  
\end{equation}
and
\begin{equation*}
	H_{F}^{(b)} = -\Biggl[ e^{i\frac{(c+1)\lambda_0}{c\omega}} \sum_{\beta\neq0} i^{\beta(1-c)} J_{-c\beta} \left(-\frac{\lambda_0}{\omega}\right) J_{\beta} \left(-\frac{\lambda_0}{c\omega}\right) \sum_{j} \tilde{\sigma}_{j}^{+} 
	  + e^{-i\frac{(c+1)\lambda_0}{c\omega}} \sum_{\beta\neq0} i^{\beta(1-c)} J_{-c\beta} \left(\frac{\lambda_0}{\omega}\right) J_{\beta} \left(\frac{\lambda_0}{c\omega}\right) \sum_{j} \tilde{\sigma}_{j}^{-} \Biggr].
\end{equation*}
We can use the Eq.~\eqref{eq:HF1} to predict the key features of the actual dynamics. For example, a substitution of $c=3$ in Eq.~\eqref{eq:HF1} yields
\begin{flalign}
	H_{F}|_{c=3} & = -\Omega \biggl[J_{0}(\frac{\lambda_0}{\omega})J_{0}(\frac{\lambda_0}{3 \omega})+\sum_{n}(-1)^{n} J_{-3n}(\frac{\lambda_0}{\omega})J_{n}(\frac{\lambda_0}{3 \omega})\biggl]\biggl[ \cos\biggl( \frac{4\lambda_0}{3\omega}\biggr) \sum_{j} P_{j-1} \sigma^x_j P_{j+1}   \nonumber\\ 
	& \quad- \sin\biggl( \frac{4\lambda_0}{3\omega}\biggr) \sum_{j} P_{j-1} \sigma^y_j P_{j+1} \biggl].
\end{flalign}
We regroup the coefficients of $\Tilde{\sigma}^{x}$ and $\Tilde{\sigma}^{y}$ to express the effective Hamiltonian as
\begin{equation*}\label{eq:HFc3}
	H_{F}|_{c=3}  = C1 \sum_{j}P_{j-1} \sigma^x_j P_{j+1} + C2 \sum_{j}P_{j-1} \sigma^y_j P_{j+1}.
\end{equation*}
We note that the coefficients $C_1$ and $C_2$ are infinite series that involve products of $J_n(1/\omega)$. Hence, we must consider its convergence at low values of $\omega$. To examine the convergence behaviour, we show in Fig.~\ref{fig:convergec3} the plots of $C_1$ and $C_2$ by truncating the series at different number of terms, $N$. We note that the series is symmetric under $n \leftrightarrow -n$. The value of $\omega$ for which $C_1=C_2 \approx 0$ indicates ergodic dynamics. We observe that for $\omega \geq 2$, there are no simultaneous intersection of $C_1$ and $C_2$ with the horizontal axis, irrespective of $N$. This indicates a non-zero PXP contribution, and as a consequence, scar-dominance, which is in agreement with the actual dynamics. However, the role of $N$ becomes significant for $\omega < 2$, where $C_1$ and $C_2$ simultaneously vanish around $\omega \approx 1.15$, $1.05$ and  $0.75$ for $N=4$, $6$, and $8$, respectively. Therefore, the Floquet perturbation theory (FPT) may not be able to accurately predict the very low-frequency dynamics. However, FPT is very useful for the intermediate and high-frequency regimes. In Fig.~\ref{fig:convergec4and5N8} (a) and (b) we show the behaviour of coefficients for $c=4$ and $5$, respectively, at $N=8$. In a similar fashion, we can show the properties of $H_F$ for other $c$.

\subsection*{Quasi-periodic drive}

To model a quasi-periodic drive with $c=\alpha=(1+\sqrt{5}/2)$, we use the FPT and a rational approximation of $\alpha$. For $\alpha=8/5$, the periodicity of the drive is $T=10\pi/\omega$. We obtain
\begin{equation} \nonumber
	(H_{F})_{n\pm 1} = -\frac{\Omega}{T}\int_{0}^{T} dt \sum_{\alpha,\beta} i^{\alpha+\beta} J_{\alpha}(\mp \lambda_{0}/\omega) J_{\beta}(\mp \lambda_{0}/c\omega)\times e^{\pm i\frac{13\lambda_0}{8\omega}} e^{i\omega(\alpha+8\beta/5)t}.
\end{equation}
For the integral to be non-zero, we must have $\alpha + 8\beta/5=0$. This yields 
\begin{equation}
	H_{F} = H_{F}^{(a)}+H_{F}^{(b)},
\end{equation}
where
\begin{equation}\nonumber
	H_{F}^{(a)}  = -\Omega J_{0} \left(\frac{\lambda_0}{\omega}\right) J_{0} (\frac{8\lambda_0}{5\omega}) [\cos  \frac{13\lambda_0}{8\omega}  \sum_{j} \tilde{\sigma}_{j}^{x}  
- \sin \frac{13\lambda_0}{8\omega} \sum_{j} \tilde{\sigma}_{j}^{y}], 
\end{equation}
\begin{align}	\nonumber
		H_{F}^{(b)} &= -\Omega \Bigl[ e^{i\frac{13\lambda_0}{8\omega}} \sum_{n=\pm 5, \pm 10, \dots} i^{-3n/5} J_{\frac{-8n}{5}}(-\frac{\lambda_0}{\omega}) J_{n}(-\frac{8\lambda_0}{5\omega}) \sum_{j} \tilde{\sigma}_{j}^{+} \\
	& \quad + e^{-i\frac{13\lambda_0}{8\omega}} \sum_{n=\pm 5, \pm 10, \dots} i^{-3n/5} J_{-\frac{8n}{5}}(\frac{\lambda_0}{\omega}) J_{n}(\frac{8\lambda_0}{5\omega}) \sum_{j} \tilde{\sigma}_{j}^{-} \Bigr]. \nonumber
\end{align}

The approximation of the quasi-periodic drive becomes better for $\alpha=13/8$, for which the drive period increases to $16\pi/\omega$. We obtain
\begin{equation}\nonumber
	H_{F}^{(a)}  = -\Omega J_{0} \left(\frac{\lambda_0}{\omega}\right) J_{0} (\frac{8\lambda_0}{13\omega}) [ \cos \left( \frac{21\lambda_0}{13\omega} \right) \sum_{j} \tilde{\sigma}_{j}^{x} 
- \sin ( \frac{21\lambda_0}{13\omega}) \sum_{j} \tilde{\sigma}_{j}^{y}] 
\end{equation}	
and 
\begin{align}	\nonumber
	H_{F}^{(b)} &= -\Omega \Bigl[ e^{i\frac{21\lambda_0}{13\omega}} \sum_{n=\pm 8, \pm 16, \dots} i^{-5n/8} J_{-\frac{13}{8}}(-\frac{\lambda_0}{\omega}) J_{n}(-\frac{8\lambda_0}{13\omega}) \sum_{j} \tilde{\sigma}_{j}^{+} \\
 & \quad	+ e^{-i\frac{21\lambda_0}{13\omega}} \sum_{n=\pm 8, \pm 16, \dots} i^{-5n/8} J_{-\frac{13n}{8}}(\lambda_0/\omega) J_{n}(\frac{8\lambda_0}{13\omega}) \sum_{j} \tilde{\sigma}_{j}^{-} \Bigr]. \nonumber
\end{align}

\begin{figure}[h!]
	\centering
	\includegraphics[width=0.95\linewidth]{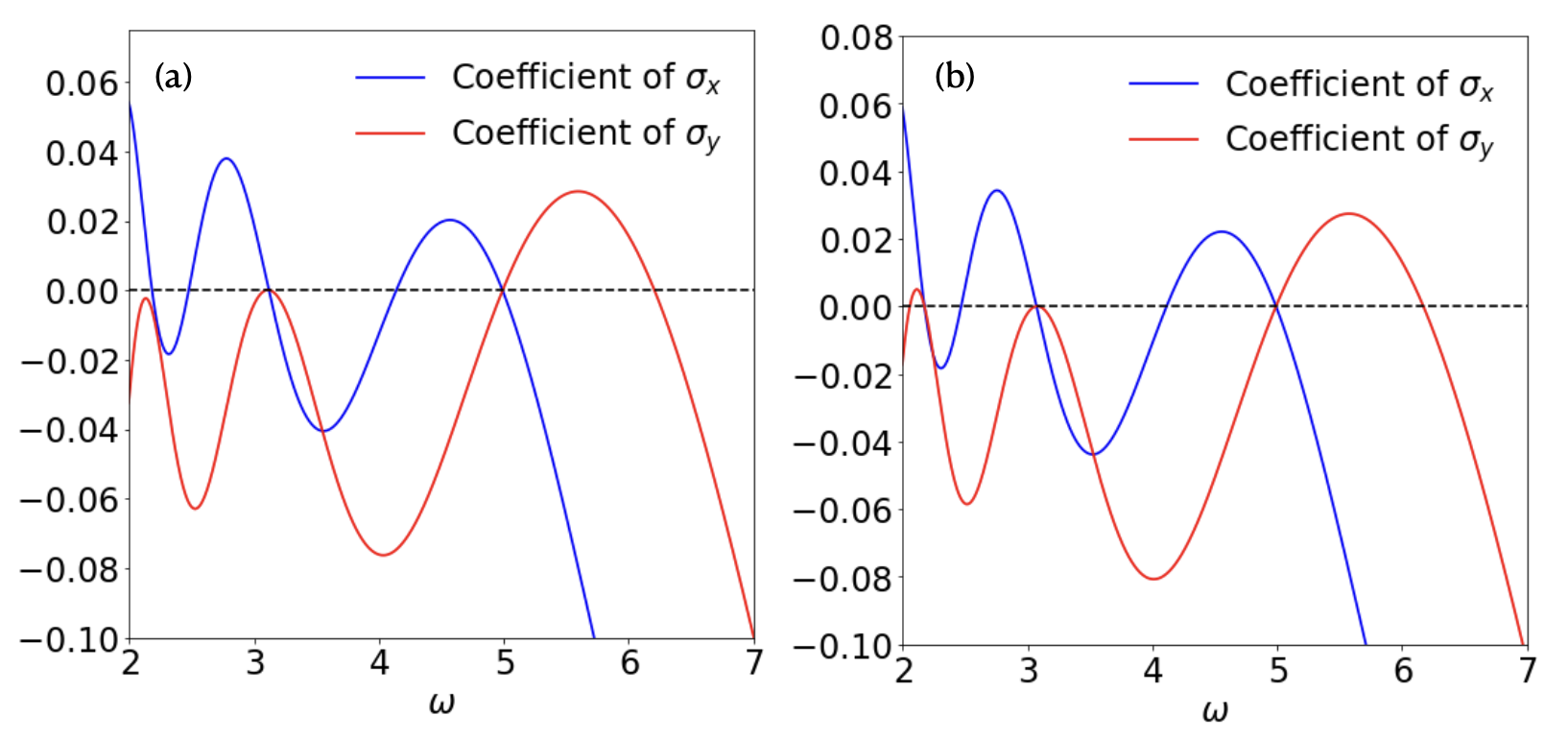}
	\caption{Coefficients of $\Tilde{\sigma}_{j}^{x}$ and $\Tilde{\sigma}_{j}^{y}$ when approximating a quasi-periodic drive with (a) $c = 8/5$ and (b) $c = 13/8$. In the former case, ergodicity is expected around $\omega \approx 3.15$ and $\omega \approx 5$, whereas in the latter, the corresponding values are $\omega \approx 2.25$, $3.15$, and $5$. In both plots, $N = 8$.}
	 \label{fig:qp8y5and13y8N8}
\end{figure}

We remark that even for small fractional values of $c$, $H_F$ contains higher order Bessel functions; thereby, distinguishing it from the case when $c$ is a small integer. Figure~\ref{fig:qp8y5and13y8N8} (a) and (b) show the behaviour of the coefficients for $c=8/5$ and $13/8$, respectively, at $N=8$. We observe that the PXP term vanishes around $\omega \approx 3.15$ and $5$ for $c=8/5$, whereas at $\omega \approx 2.25$, $3.15$ and $5$ for $c=13/8$. Also, for $\omega>5$, no such $\omega$ exists for which the PXP term can vanish, this suggests the presence of scar-induced oscillations. Note that these observations closely align with the corresponding behaviour of the fidelity $\langle F \rangle$. 

Finally, for $c=0$ and $1$, the driving protocol simplifies to a single-frequency periodic drive with drive-amplitudes $\lambda_0$ and $2\lambda_0$, respectively. For $c=1$, the dynamics is governed by the zeros of $J_0(2\lambda_0/\omega)$, i.e., the PXP term vanishes around $\omega \approx 1.61$, $2.03$, $2.77$, $4.35$, $9.98$, \dots. Furthermore, dynamical properties for drives with $c=1/m$, where $m$ is a real number, can be predicted due to the symmetry of the driving protocol with respect to the interchange of $\omega_1$ and $\omega_2$. Also, it is possible to demonstrate that the critical frequencies for $c=m$, where thermalization occurs are $m$ times the critical frequencies corresponding to $c=1/m$. As a consequence, for $m \gg 1$ (small $c$), the non-monotonic transitions can persist up to much larger $\omega$.

\subsection*{B. Dynamics of the polarized state}
\begin{figure}[h!]
	\centering
	\includegraphics[width=0.95\linewidth]{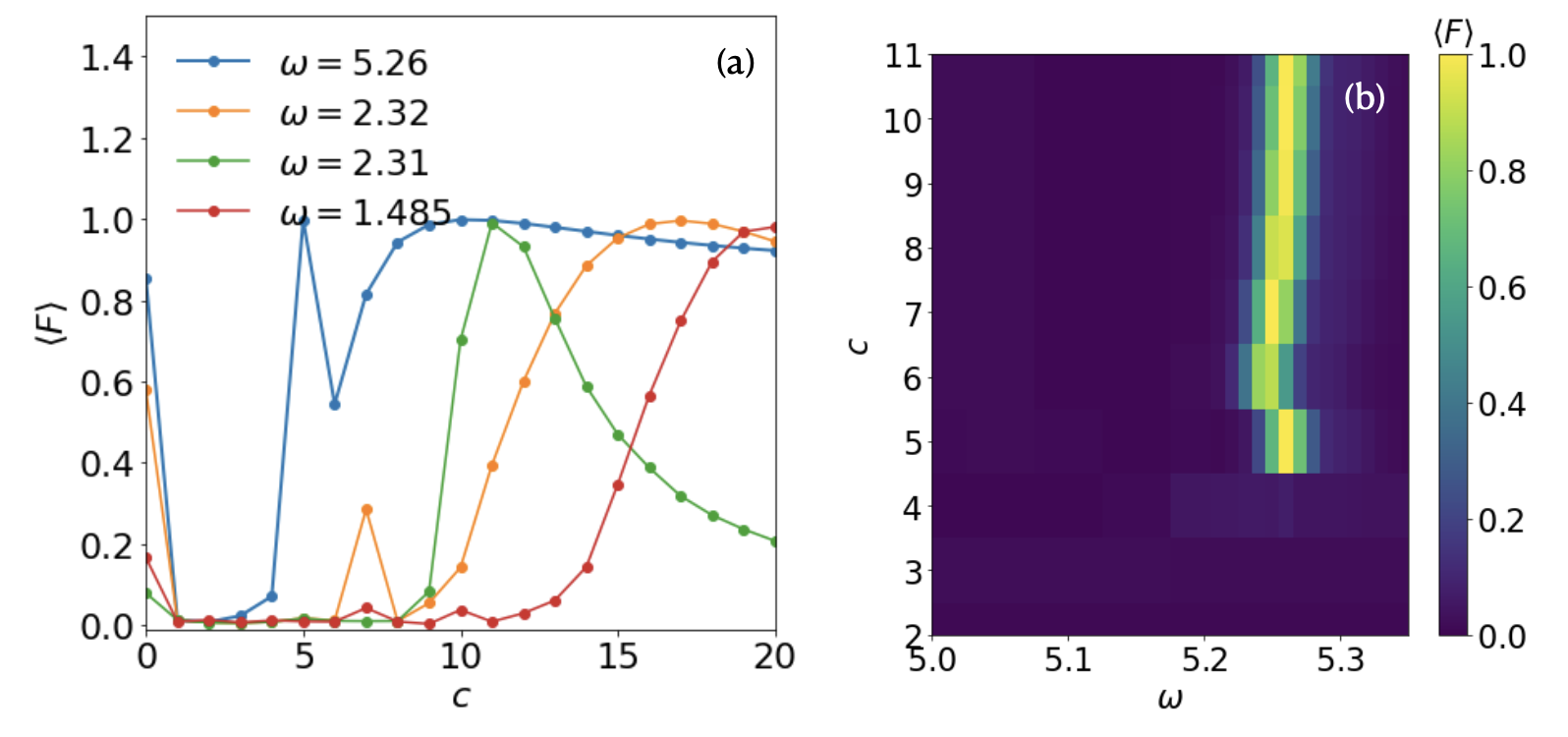}
	\caption{Fate of dynamical freezing and evolution of the polarized state under a two-frequency drive. (a) $\langle F \rangle$ vs $c$ near the three roots of $J_{0}(\lambda_0/\omega)$. Dynamical freezing is absent for $c = 2, 3, 4$, with the transition point shifting to higher $c$ as $\omega$ decreases. (b) Density plot of $\langle F \rangle$ in the $\omega$–$c$ plane, indicating the onset of dynamical freezing at $c = 5$ near $\omega \approx 5.26$. In both plots, $\lambda_0 = 12$.}
	 \label{fig:freez}
\end{figure}

In the presence of periodic or quasi-periodic drive, the dynamics started from the polarized state $\ket{O}=\ket{\downarrow\downarrow\downarrow}$ exhibits ergodicity for $\omega \gg \lambda_0$. However, for $\omega \ll \lambda_0$, there exist certain frequencies at which $U \to I$, suggesting the presence of many-body freezing~\cite{haldar2021dynamical}. Interestingly, these freezing points approximately correspond to the roots of $J_{0}(\lambda_0/\omega)=0$~\cite{dutta2025prethermalization}. To further examine this, we show in Fig.~\ref{fig:freez} (a) the behaviour of $\langle F \rangle$ vs $c$ for different values of $\omega$ for which $J_0(\lambda_0/\omega)=0$. We find that for $\omega=5.26$ ($\lambda_0/\omega \approx 2.28$), dynamical freezing occurs at $c=5$ and for $c \geq 7$. Moreover, as the frequency decreases ($\lambda_0/\omega \approx 5.19$, $8.08$, \dots), dynamical freezing appears to vanish for smaller values of $c$. For small values of $c$, the effective Hamiltonian $H_F$, when higher order terms are incorporated, should be able to describe the describe the dynamics starting from a polarized state. However, the calculations involved are non-trivial and we leave it for future explorations.

\subsection*{C. Effect of perturbations}

\begin{figure}[h!]
	\includegraphics[width=0.95\linewidth]{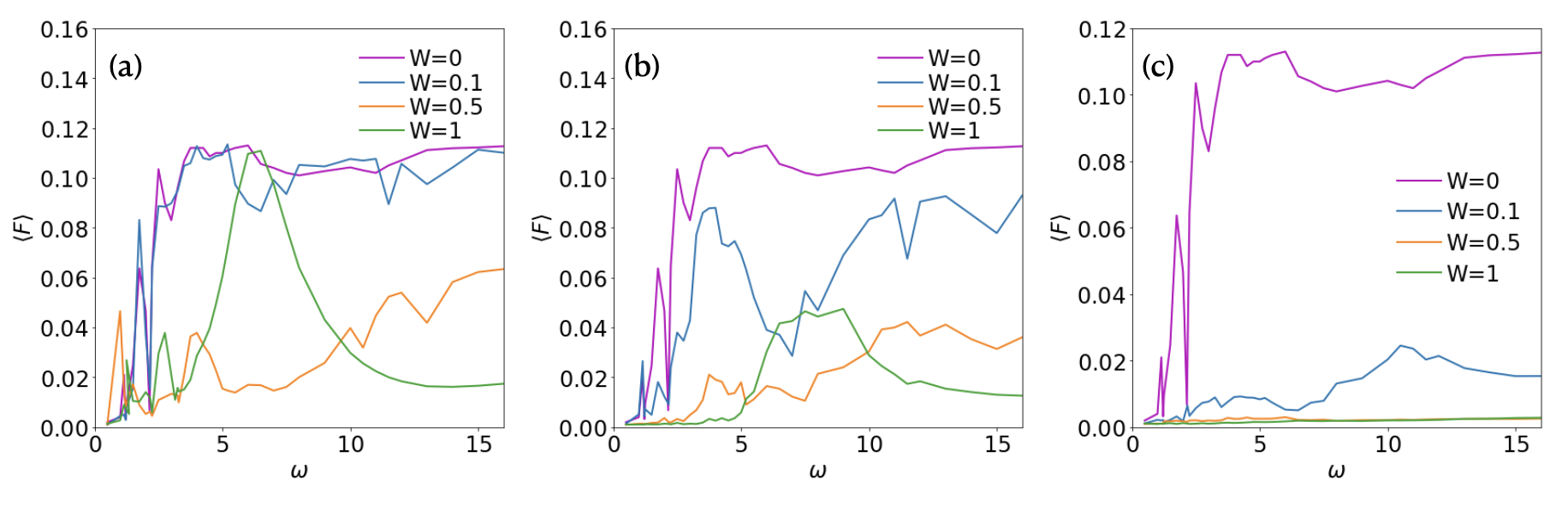}
	\caption{Effect of different perturbations: (a) $H_{p}(1)$, (b) $H_{p}(2)$, and (c) $H_{p}(3)$ for $c = 2$. In the high-frequency regime ($\omega > 10$), $\langle F \rangle$ decreases with $W$ in all cases, whereas for $\omega \leq 10$, $\langle F \rangle$ exhibits non-monotonic behavior with W, particularly for $H_{p}(1)$ and $H_{p}(2)$. In contrast, for $H_{p}(3)$, even a small $W$ can significantly reduce the scarring features. In all plots, $\lambda_0$ is fixed at 12.}
	\label{fig:Hpc2}
\end{figure}

\begin{figure}[h!]
	\includegraphics[width=0.95\linewidth]{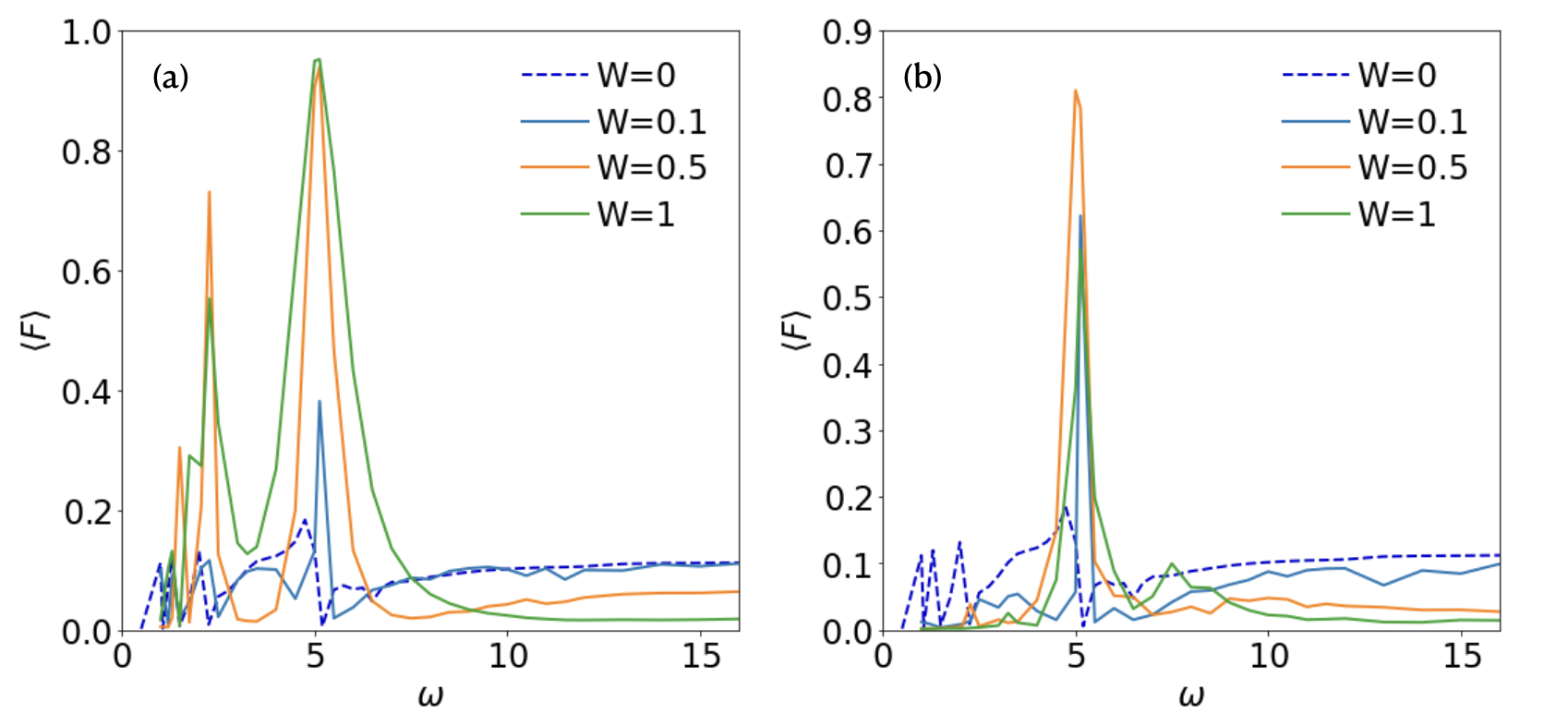}
	\caption{Effect of different perturbations: (a) $H_{p}(1)$, (b) $H_{p}(2)$, and (c) $H_{p}(3)$ for $c = 20$. In the low-frequency regime ($\omega \leq 10$), $\langle F \rangle$ displays pronounced non-monotonic dependence on $W$, with dynamical freezing emerging at specific $\omega$ for $H_{p}(1)$ and $H_{p}(2)$. Other features resemble those in Fig.~\ref{fig:Hpc2}. In all the plots, $\lambda_0$ is fixed at 12.}
	\label{fig:Hpc20}
\end{figure}

Next, we examine the influence of different perturbations on the dynamics of the PXP model under two-frequency drive. The perturbations can either annihilate or enhance the scar-induced oscillations~\cite{hudomal2022driving}. We consider the following classes of perturbations:
\begin{flalign}
	H_p(1) & = H(t) + W\sum_{i} n_{i}, \label{eq:Hp1}\\
	H_p(2) & = H(t) + W\sum_{i} P_{i-1}(\sigma_{i}^{+}\sigma_{i+1}^{-}+\sigma_{i}^{-}\sigma_{i+1}^{+})P_{i+2}, \label{eq:Hp2}\\ 
	H_p(3) & = H(t) + W\sum_{i} P_{i-1}\sigma_{i}^{x}P_{i+1}\sigma_{i+2}^{x}P_{i+3}, \label{eq:Hp3}
\end{flalign}
where Eq.~\eqref{eq:Hp1} corresponds to the addition of an onsite uniform chemical potential term, Eq.~\eqref{eq:Hp2} and Eq.~\eqref{eq:Hp3} represent constrained nearest-neighbor hopping and next-nearest neighbor flips, respectively. We choose $c=2$ and illustrate the behavior of $\langle F \rangle$ vs $\omega$ for $H_p(1)$, $H_p(2)$ and $H_p(3)$ in  Fig.~\ref{fig:Hpc2} (a), (b) and (c), respectively. We observe that in the high-frequency regime, regardless of the nature of the perturbation, $\langle F \rangle$ decreases consistently with the strength of the perturbation. The next-nearest neighbor perturbation is most effective in destroying the scar-induced oscillations, wherein for $W=0.1$ the maximum value of $\langle F \rangle$ drops to approximately $0.02$. In contrast to this, for the nearest neighbor perturbation, $\langle F \rangle$ exhibits a non-monotonic variation with $W$. We find that over the range $5 \leq \omega \leq 10$, $\langle F \rangle$ for $W=1$ exceeds the values observed for $W=0.5$. When the frequency is tuned below $5$, $\langle F \rangle$ decreases with increasing $W$. Similarly, the presence of a uniform chemical potential leads to a significant non-monotonicity in the dynamics over the frequency ranges $2 \leq \omega < 3.5$ and $4.5 \leq \omega \leq 10$; for $W=1$, $\langle F \rangle$ peaks around $\omega \approx 6.5$. In order to gain further insights,  we illustrate the behavior of $\langle F \rangle$ vs $\omega$ for different perturbation strengths in Fig.~\ref{fig:Hpc20}; we keep $c=20$ fixed. Interestingly, the dynamics under $H_p(1)$ exhibits $\langle F \rangle \approx 1$ near $\omega \approx 5.13$, which indicates the presence of a dynamical freezing. We observe similar signatures for the case of $H_p(2)$; however, in the high-frequency regime, the dynamics resembles that observed for $c=2$.

\begin{figure}[h!]
	\centering
	\includegraphics[width=0.35\linewidth]{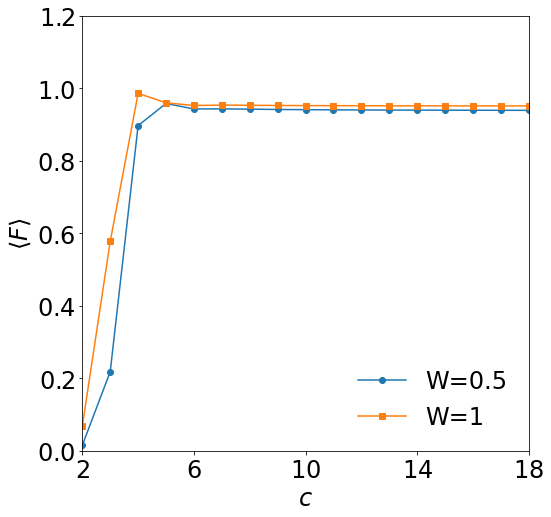}
	\caption{$\langle F \rangle$ as a function of $c$ for $H_{p}(1)$. At large perturbation strengths ($W=0.5$, $W=1$), signatures of dynamical freezing are observed for $c\geq 4$.}
	\label{fig:dynfreezc}
\end{figure}

Finally, to study the effect of $c$ on the existence of dynamical freezing, we show the behavior of $\langle F \rangle$ vs $c$ for $H_p(1)$ near $\omega \approx 5.13$ in Fig.~\ref{fig:dynfreezc}. We observe that for $c \geq 4$, $\langle F \rangle$ remains close to unity. Thus, when the ratio of the driving frequencies is large, the resulting dynamics is similar to the case of a single-frequency drive, the latter exhibits dynamical freezing. Therefore, our analysis demonstrate the above non-ergodic phenomenon persists in a two-frequency driving protocol, even when the separation the two time scales is not particularly large.

\subsection*{D. System-size scaling}
\begin{figure}[h!]
	\centering
	\includegraphics[width=0.95\linewidth]{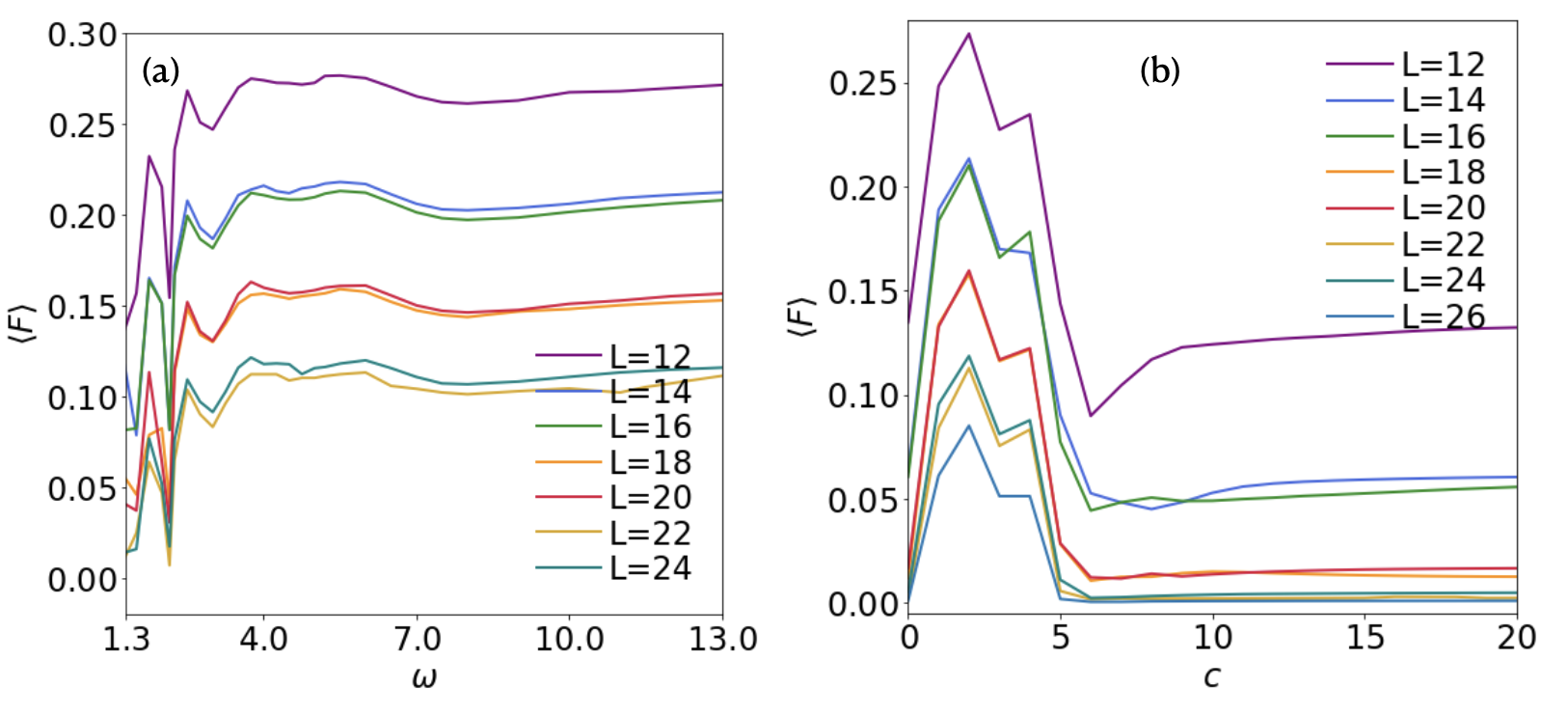}
	\caption{System-size dependence in the two-frequency driven PXP model. (a) $\langle F \rangle$ as a function of $\omega$, showing a decreasing trend with increasing $L$. However, scarring remains significant up to moderately large system sizes ($L \leq 20$), even at low frequency. (b) $\langle F \rangle$ as a function of $c$ around $\omega \approx 5.15$ for different $L$. The (non-)ergodic features at $c = 0$ and $c \geq 5$ appear to be the same for all $L$. In both plots, $\lambda_0$ is fixed at 12.}
	\label{fig:syssize}
\end{figure}

We conclude with a discussion of the effect of system size ($L$) on the dynamics. In Fig.~\ref{fig:syssize} (a) we show the time-averaged fidelity $\langle F \rangle$ vs $\omega$ for different values of system sizes at $c=2$. We find that as $L$ increases, $\langle F \rangle$ decreases, indicating a reduction in the amplitude of scar-induced oscillations. However, we find that the ergodicity is absent for $L \leq 20$ when $\omega \geq 1.5$. Thus, fidelity revivals in a two-frequency driven PXP model can persist up to larger system sizes. Figure~\ref{fig:syssize} (b)
shows the behavior of $\langle F \rangle$ vs $c$ around $\omega = 5.15$ for different values of $L$. A clear transition from non-ergodic to ergodic behavior sets in at $c=5$ for $L \geq 20$. Note that these transitions are also observed for $L<20$, but are not significant enough to completely suppress the scarred oscillations.

﻿

﻿
﻿
\end{document}